\documentstyle[12pt]{article}
\begin{document}
\begin{center}
{\Large Strong coupling in the Kondo problem in the \\
low-temperature region}
\end{center}
\bigskip
\centerline{\bf Yu.N. Ovchinnikov and A.M. Dyugaev}

\begin{center}
{\small L.D.~Landau Institute for Theoretical Physics,\\ 
117940 Moscow, Russia}
\end{center}



\vspace{1cm}
\section{Abstract}

The magnetic field dependence of the average  spin of a 
localized electron coupled to conduction electrons with an 
antiferromagnetic exchange interaction is found for the ground 
state. In the magnetic field range $\mu H\sim 0.5 T_c$ 
($T_c$  is the Kondo temperature) there is an inflection
point, and in the  
strong magnetic field range $\mu H\gg T_c$, the correction to the 
average spin is proportional to $(T_c/\mu H)^2$. In zero magnetic field, the 
interaction with conduction electrons also leads  to the 
splitting of doubly degenerate spin impurity states. 

\vspace{0.5cm}
\section{Introduction}

In the low-temperature and weak magnetic field region, 
even a weak interaction of magnetic impurities with a degenerate electron 
gas becomes  strong$^{1-3}$. In this region, perturbation
theory is violated. Two scenarios are possible in  such a situation. 
First, an assumption can be made that in the low-temperature region,
an  increase
in the magnetic field takes the system out of a strongly coupled state and 
into the region of applicability of perturbation theory. 
This nonobvious  
conjecture was used  in Bethe's ansatz method in the problem under 
consideration. As the result, in a strong magnetic field  $\mu_eH\gg T_c$
($T_c$  is the Kondo temperature), the  correction to the mean spin impurity 
value has  logarithmic behavior$^3$,
$\langle S_z \rangle =\frac{1}{2} \Bigl ( 1-\frac{1}{2\ln (\mu_eH/T_c)} 
\Bigr )$. Such  spin dependence 
of the magnetic field   value
is too slow, and is inconsistent with  
the  experimental data$^4$, which yields  
power-like behavior. The level of spin satiation in the magnetic field 
in  Ref.4 (Fig. 8) can be reached according to the expression
given above only at the magnetic field value $H\approx 50$ T, instead 
of the experimental value of 6 T. 

The second scenario is connected with the assumption that an increase only 
in the magnetic field value does not move  the system from a strongly coupled 
state to a weak the perturbed state. The second conjecture is 
supported by the fact that the correction to the wave function 
of a system consisting of  magnetic impurity plus degenerate 
Fermi gas, in some state 
with low energy, contains corrections of two types 
obtained with the help of perturbation 
theory. The norm of one of them decreases in an increasing magnetic 
field, whereas the norm of the other is divergent in the limit 
$T\to 0$ for a finite magnetic  field. Consideration of the 
norm of states in the problem involved is very useful, 
because it contains direct information about the average value of 
magnetic spin. 

Below we consider in detail the second conjecture and confirm it. In 
the low-temperature region $(T \ll T_c)$, the average  spin of 
magnetic impurities is found for an arbitrary value of the external
magnetic field. States for both signs of interaction constant 
are investigated. The strong coupled state arises in both cases, 
but the magnetic field dependence of the average value of spin is 
substantially  different. The definition of Kondo temperature $T_c$               
is also slightly different for different signs of the interaction 
constant. 

\vspace{0.5cm}
\section{The model}

We will suppose that the interaction of magnetic impurity with the 
Fermi sea of 
electrons has an exchange nature. Then the Hamiltonian $\hat H$ of the system 
under consideration can be taken in the form
\begin{equation}
\hat H=\hat H_0+\int d^3r_1d^3r_2V(r_1-r_2)
\chi^{+}_{\alpha}(r_1)\varphi^{+}_{\beta}(r_2)
\chi_{\beta}(r_2)\varphi_{\alpha}(r_1)
\end{equation}
\begin{displaymath}
-\frac{\mu H}{2} \int \Bigl ( \varphi^{+}_{\uparrow}(r_1)
\varphi_{\uparrow}(r_1)-
\varphi^{+}_{\downarrow}(r_1)\varphi_{\downarrow}(r_1) \Bigr ) d^3r_1.
\end{displaymath}

In Eq. (1), operators $\varphi^{+}_{\beta},~\chi^{+}_{\alpha}$ 
are creation operators of an electron in a localized state on a magnetic 
impurity and in the continuum spectrum respectively. For simplicity, 
we consider the case with one unpaired electron in the localized 
state (spin 1/2). The first  term in Eq. (1) describes the degenerate
electron gas in some external field that leads to creation of 
one localized state. The spin interaction of electrons in the continuum 
spectrum with magnetic field leads only to small renormalization 
of the magnetic moment of a localized electron, and a small shift in the
kinetic energy of electrons with spin up and down in such a 
way that they have the same value of chemical potential (no gap for 
transfer of electron with spin flip over the Fermi level). For 
this reason we omit this term in Hamiltonian (1). The last 
term gives the interaction energy of a localized electron with the
magnetic field. 

Consider now the limiting case as $T\to 0$ and $H$  finite. We 
search for the lowest-energy eigenfunction $|\psi\rangle$ of 
Hamiltonian (1) in Fock space in the form
\begin{equation}
|\psi\rangle =|10;11;11;..\rangle+
\sum C^{2L-1}_{2K}|01;\stackrel{2K}{10};\stackrel{2L-1}{10}\rangle+
\sum C^{2L-1}_{2K-1}|10;\stackrel{2K-1}{01};\stackrel{2L-1}{10};\rangle 
\end{equation}
\begin{displaymath}
+\sum C^{2L}_{2K}|10;\stackrel{2K}{10};\stackrel{2L}{01}\rangle +
\sum_{K_1<K}C^{2L_1~2L-1}_{2K_1~2K}
\hat N|01;\stackrel{2K_1}{10};\stackrel{2K}{10};
\stackrel{2L_1}{01};\stackrel{2L-1}{10}\rangle
\end{displaymath}
\begin{displaymath}
+\sum C^{2L_1;2L-1}_{2K_1-1;2K}
\hat N|10;\stackrel{2K_1-1}{01};\stackrel{2K}{10};
\stackrel{2L_1}{01};\stackrel{2L-1}{10}\rangle+
\sum_{L_1<L}C^{2L_1-1;2L-1}_{2K_1;2K-1}
\hat N|01;\stackrel{2K_1}{10};\stackrel{2K-1}{01};
\stackrel{2L_1-1}{10};\stackrel{2L-1}{10}\rangle
\end{displaymath}
\begin{displaymath}
+\sum_{K_1<K;L_1<L} 
C^{2L_1-1;2L-1}_{2K_1-1;2K-1}
\hat N|10;\stackrel{2K_1-1}{01};\stackrel{2K-1}{01};
\stackrel{2L_1-1}{10};\stackrel{2L-1}{10} \rangle
\end{displaymath}
\begin{displaymath}
+\sum_{K_1<K;L_1<L}
C^{2L_1;2L}_{2K_1;2K}
\hat N|10;\stackrel{2K_1}{10};\stackrel{2K}{10};
\stackrel{2L_1}{01};\stackrel{2L}{01}\rangle+...
\end{displaymath}
In Eq. (2), all single-particle states (solutions of Eq. (1) for one 
particle) are ordered and numbered. Indexes $K,L$ label states
under and over the Fermi surface. Each box has two places. The 
first one means a 
state with spin up, and the second with spin down. 
As an example,  the state
$|\stackrel{2K}{10};\stackrel{2L}{01}\rangle$ means that the state $2K$
(spin down) under the Fermi surface is empty and the state $2L$ 
(spin down) over the Fermi surface is filled. The first cell is
always  reserved for an electron in a localized state. The first term in 
Eq. (2) gives the ground state of Hamiltonian (1) without interaction
$(V(r)=0)$. The number of upper (or lower) indexes in 
$C^{\cdot\cdot\cdot}_{\cdot\cdot\cdot}$ gives the 
number of excited pairs. For $P$ excited pairs, there are $2P+1$ 
different symbols 
$C^{\cdot\cdot\cdot}_{\cdot\cdot\cdot}$. 
Operator 
$\hat N$ is the ordering operator, and each rearrangement of two 
neighboring filled states gives a factor (-). 
In Eq. (2) in each box below Fermi surface, only one place can 
be empty and above the Fermi surface in each box, only one place can be 
filled. 

The equation for the wave function $|\psi\rangle$ is 
\begin{equation}
|\hat H\psi\rangle =E|\psi\rangle,
\end{equation}
where $E$ is the energy of the state.

Inserting expression (2) for the wave function $|\psi\rangle$ into 
Eq. (3), we obtain a set of linear equations for the quantities 
$C^{\cdot\cdot\cdot}_{\cdot\cdot\cdot}$. Due to the structure of 
Hamultonian (1), each quantity 
$C^{\cdot\cdot\cdot}_{\cdot\cdot\cdot}$ order of $P$ is coupled 
only with quantities $C^{\cdot\cdot\cdot}_{\cdot\cdot\cdot}$ 
order of $P,P\pm 1$. From the first equation of this system, we 
obtain the energy of the state,
\begin{equation}
E=E_0-\mu H/2-\delta E,
\end{equation}
\begin{displaymath}
\delta E=\sum \Bigl (
I^{2L-1^*}_{2K-1}C^{2L-1}_{2K-1}-
I^{2L-1^*}_{2K}C^{2L-1}_{2K} \Bigr ),
\end{displaymath}
where $E_0$ is the energy of the ground state without interaction. For 
convenience, we leave the magnetic energy of the localized state 
out of the  
correction term $\delta E$. The quantities $I^{\cdot}_{\cdot}$ 
in Eq. (4) are the transition 
matrix elements. As an example, we have
\begin{equation}
I^{2L-1}_{2K}=
\int d^3r_1d^3r_2\chi^{*}_{\uparrow}(r_1)
\varphi^{*}_{\downarrow}(r_2)\varphi_{\uparrow}(r_1)
\chi_{\downarrow}(r_2)V(r_1-r_2).
\end{equation}

The Hamiltonian (1) possesses deep symmetry properties. To 
see some of these, we will keep indexes on $I$
that indicate energy and spin in the initial and 
final states. The next three equations for the quantities 
$C^{\cdot}_{\cdot}$  are 
\begin{equation}
-I^{2L-1}_{2K}+
\sum C^{2L-1}_{2K_1}I^{2K_1}_{2K}-
\sum C^{2L-1}_{2K_1-1}I^{2K_1-1}_{2K}-
\sum C^{2L_1}_{2K}I^{2L-1}_{2L_1}
\end{equation}
\begin{displaymath}
+(\mu H+\varepsilon_L-\varepsilon_K-\delta E)C^{2L-1}_{2K}+
\sum_{K_1<K}C^{2L_1;2L-1}_{2K_1;2K}I^{2K_1}_{2L_1}
\end{displaymath}
\begin{displaymath}
-\sum_{K<K_1}C^{2L_1;2L-1}_{2K;2K_1}I^{2K_1}_{2L_1}-
\sum C^{2L_1;2L-1}_{2K_1-1;2K}I^{2K_1-1}_{2L_1}=0,
\end{displaymath}
\begin{displaymath}
I^{2L-1}_{2K-1}-
\sum I^{2K_1}_{2K-1}C^{2L-1}_{2K_1}+
\sum C^{2L-1}_{2K_1-1}I^{2K_1-1}_{2K-1}-
\sum I^{2L-1}_{2L_1-1}C^{2L_1-1}_{2K-1}
\end{displaymath}
\begin{displaymath}
+(\varepsilon_L-\varepsilon_K-\delta E)C^{2L-1}_{2K-1}+
\sum_{L_1<L}C^{2L_1-1;2L-1}_{2K_1;2K-1}I^{2K_1}_{2L_1-1}
\end{displaymath}
\begin{displaymath}
-\sum_{L<L_1}C^{2L-1;2L_1-1}_{2K_1;2K-1}I^{2K_1}_{2L_1-1}+
\sum_{K<K_1;L_1<L} C^{2L_1-1;2L-1}_{2K-1;2K_1-1}I^{2K_1-1}_{2L_1-1}-
\sum_{K_1<K;L_1<L}C^{2L_1-1;2L-1}_{2K_1-1;2K-1}I^{2K_1-1}_{2L_1-1}
\end{displaymath}
\begin{displaymath}
-\sum_{L<L_1;K<K_1}C^{2L-1;2L_1-1}_{2K-1;2K_1-1}I^{2K_1-1}_{2L_1-1}+
\sum_{K_1<K;L<L_1} C^{2L-1;2L_1-1}_{2K_1-1;2K-1}I^{2K_1-1}_{2L_1-1}=0,
\end{displaymath}
\begin{displaymath}
-\sum I^{2L}_{2L_1-1}C^{2L_1-1}_{2K}+
(\varepsilon_L-\varepsilon_K-\delta E)C^{2L}_{2K}+
\sum_{K<K_1}C^{2L;2L_1-1}_{2K;2K_1}I^{2K_1}_{2L_1-1}
\end{displaymath}
\begin{displaymath}
-\sum_{K_1<K}C^{2L;2L_1-1}_{2K_1;2K}I^{2K_1}_{2L_1-1}+
\sum C^{2L;2L_1-1}_{2K_1-1;2K}I^{2K_1-1}_{2L_1-1}=0.
\end{displaymath}
In Eq. (6), the quantities $\varepsilon_{L,K}$ are the energies of single
states. As mentioned above, index $L$ means a state above the
Fermi level and index $K$ means a state below the Fermi level. The 
equations for $C^{\cdot\cdot}_{\cdot\cdot}$ are
given in Appendix A. Since the equations for  
$C^{\cdot\cdot\cdot}_{\cdot\cdot\cdot}$ have a special structure,
quantity $C^{\cdot\cdot\cdot}_{\cdot\cdot\cdot}$ order of $P$
is coupled only with quantities $C^{\cdot\cdot\cdot}_{\cdot\cdot\cdot}$ 
order of $P,P\pm 1$, it is possible to leave quantities 
$C^{\cdot\cdot\cdot}_{\cdot\cdot\cdot}$ order of $P\geq 2$ out of Eqs.
(6). As the result, we obtain three equations for the quantities 
$C^{2L-1}_{2K},C^{2L-1}_{2K-1}$ and $C^{2L}_{2K}$. They have the 
following form (from Appendix A):
\begin{equation}
-I^{2L-1}_{2K}+\sum C^{2L-1}_{2K_1}I^{2K_1}_{2K}-
\sum C^{2L-1}_{2K_1-1}I^{2K_1-1}_{2K}-
\sum C^{2L_1}_{2K}I^{2L-1}_{2L_1}
\end{equation}
\begin{displaymath}
+\Bigl ( \mu H+\varepsilon_L-\varepsilon_K-\delta E-\Sigma^{(1)}_{(K,L)} \Bigr )
C^{2L-1}_{2K}=A_1\Bigl ( C^{2L-1}_{2K};C^{2L-1}_{2K-1};C^{2L}_{2K} \Bigr ),
\end{displaymath}
\begin{displaymath}
I^{2L-1}_{2K-1}-\sum C^{2L-1}_{2K_1}I^{2K_1}_{2K-1}+
\sum C^{2L-1}_{2K_1-1}I^{2K_1-1}_{2K-1}-
\sum C^{2L_1-1}_{2K-1}I^{2L-1}_{2L_1-1}
\end{displaymath}
\begin{displaymath}
+\Bigl ( \varepsilon_L-\varepsilon_K-\delta E-\Sigma_{(K,L)} \Bigr )
C^{2L-1}_{2K-1}=A_2\Bigl ( C^{2L-1}_{2K};C^{2L-1}_{2K-1};C^{2L}_{2K} \Bigr ),
\end{displaymath}
\begin{displaymath}
-\sum I^{2L}_{2L_1-1}C^{2L_1-1}_{2K}+
\Bigl ( \varepsilon_L-\varepsilon_K-\delta E-\Sigma_{(K,L)}\Bigr )
C^{2L}_{2K}=A_3\Bigl ( C^{2L-1}_{2K};C^{2L-1}_{2K-1};C^{2L}_{2K} \Bigr ).
\end{displaymath}
The linear operators $A_{1,2,3}$ do not contain terms proportional 
to the quantities $C^{2L-1}_{2K},C^{2L-1}_{2K-1},C^{2L}_{2K}$  
without integral over one of variable $K,L$ with some function 
of $K,L$. These terms form the $\Sigma^{(1)}_{(K,L)},\Sigma_{(K,L)}$
terms in Eq. (7). All off-diagonal elements of such a form 
are equal to zero. The linear operators 
$A_{1,2,3}$ also do not contain terms proportional to the 
convolution of quantities $C^{\cdot}_{\cdot}$ with $I^{\cdot}_{\cdot}$
over one of variable $K,L$ without of denominator with the same variable. 
In Appendix B, we give the expressions for quantities 
$\Sigma^{(1)}_{(K,L)},~\Sigma_{(K,L)}$ in the fourth order 
of perturbation 
theory and quantities $C^{2L-1}_{2K},~C^{2L-1}_{2K-1},~C^{2L}_{2K}$
in the third order. It is easy to check that in the fourth order 
of perturbation theory, 
\begin{equation}
-\delta E-\Sigma_{(K,L)} \Bigl |_{\mbox~ {\rm for}~ \varepsilon_K=
\varepsilon_L=\varepsilon_F} \Bigr. =0.
\end{equation} 
This equality holds in all the orders of perturbation 
theory. Below, we put 
\begin{equation}
-\delta E-\Sigma_{(K,L)} \Bigl |_{\varepsilon_K=
\varepsilon_L=\varepsilon_F} =\Delta.
\end{equation} 
In Eq. (9), $\Delta \equiv\Delta (H)$ is some function of the magnetic 
field that must be determined from  self-consistency. 
This equation is  given below. Very important properties 
follow from the normalisation of states defined by Eqs. (2) and (7). To 
simplify the investigation of Eqs. (7), we give also the 
expression for operators $A_{1,2,3}$ in the lowest order of 
perturbation theory in Appendix B. All statements made above are 
independent of the exact form of spectrum $\varepsilon_{K_1},\varepsilon_L$
and  potential $V(r)$.

\vspace{0.5cm}
\section{Wave function of the ground state}

The average electron spin $\langle S_z\rangle$ in 
a bound state at zero temperature can be found by differentiating the 
energy $\delta E$ with respect to  $\mu H$
\begin{equation}
\langle S_z\rangle =\frac{1}{2}-\frac{\partial\delta E}
{\partial\mu H}.
\end{equation}
In accordance with quantum mechanical rules, the quantity 
$\langle S_z\rangle$ in the ground state is also given by an 
expression containing only norms of the states in expansion (2):
\begin{equation}
\langle S_z\rangle=
\frac{1}{2}\Bigl \{ \Bigr. 1+
\Bigl |C^{2L-1}_{2K-1}\Bigr |^2+
\Bigl |C^{2L}_{2K}\Bigr |^2-
\Bigl |C^{2L-1}_{2K}\Bigr |^2+
\Bigl |C^{2L_1;2L-1}_{2K_1-1;2K}\Bigr |^2+
\Bigl |C^{2L_1-1;2L-1}_{2K_1-1;2K-1}\Bigr |^2
\end{equation}
\begin{displaymath}
+\Bigl |C^{2L_1;2L}_{2K_1;2K}\Bigr |^2-
\Bigl |C^{2L_1;2L-1}_{2K_1;2K}\Bigr |^2-
\Bigl |C^{2L_1-1;2L-1}_{2K_1;2K-1}\Bigr |^2+... \Bigl. \Bigr \}~
\end{displaymath}
\begin{displaymath}
\times\Bigl \{ \Bigr. 1+
\Bigl |C^{2L-1}_{2K-1}\Bigr |^2+
\Bigl |C^{2L}_{2K}\Bigr |^2+
\Bigl |C^{2L-1}_{2K}\Bigr |^2+
\Bigl |C^{2L_1;2L-1}_{2K_1-1;2K}\Bigr |^2+
\Bigl |C^{2L_1-1;2L-1}_{2K_1-1;2K-1}\Bigr |^2
\end{displaymath}
\begin{displaymath}
+\Bigl |C^{2L_1;2L}_{2K_1;2K}\Bigr |^2+
\Bigl |C^{2L_1;2L-1}_{2K_1;2K}\Bigr |^2+
\Bigl |C^{2L_1-1;2L-1}_{2K_1;2K-1}\Bigr |^2+... \Bigl. \Bigr \}^{-1}~.
\end{displaymath}

Below we use both Eqs. (10) and (11). To solve Eqs. (7) and (9),
we consider $\Delta$ as a parameter. Then the right-hand side of 
Eq. (7) can be 
taken into account in perturbation theory. In the leading 
approximation we obtain
\begin{equation}
-I^{2L-1}_{2K}+\sum C^{2L-1}_{2K_1}I^{2K_1}_{2K}-
\sum C^{2L-1}_{2K_1-1}I^{2K_1-1}_{2K}-
\sum C^{2L_1}_{2K}I^{2L-1}_{2L_1}
\end{equation}
\begin{displaymath}
+(\mu H+\varepsilon_L-\varepsilon_K+\Delta)C^{2L-1}_{2K}=0~,
\end{displaymath}
\begin{displaymath}
I^{2L-1}_{2K-1}-\sum C^{2L-1}_{2K_1}I^{2K_1}_{2K-1}+
\sum C^{2L-1}_{2K_1-1}I^{2K_1-1}_{2K-1}-
\sum C^{2L_1-1}_{2K-1}I^{2L-1}_{2L_1-1}
\end{displaymath}
\begin{displaymath}
+(\varepsilon_L-\varepsilon_K+\Delta)C^{2L-1}_{2K-1}=0~,
\end{displaymath}
\begin{displaymath}
-\sum I^{2L}_{2L_1-1}C^{2L_1-1}_{2K}+
(\varepsilon_L-\varepsilon_K+\Delta )C^{2L}_{2K}=0~.
\end{displaymath}

Below we make the usual assumptions about the energy-independent value 
of the density of states near the Fermi surface, and that the 
characteristic energy in transition matrix elements $I^{\cdot}_{\cdot}$
is also the Fermi energy $\varepsilon_F$. As a result, we can put
\begin{equation}
\sum_KI^{\cdot}_{2K}() \rightarrow g\int^{\varepsilon_F}_{0}dx(), \quad
\sum_LI^{2L}_{\cdot}\rightarrow g\int^{A\varepsilon_F}_0dy(),
\end{equation}
\begin{displaymath}
\varepsilon_L-\varepsilon_F=y; \qquad \varepsilon_F-\varepsilon_K=x~.
\end{displaymath}

In Eq. (13), $g$ is the dimensionless coupling constant. The 
potential $V(r)$ in Hamiltonian (1) is in natural units,
hence the
smallness of the coupling constant $g$ is connected only to the
small radius of bound state.

Due to the energy independence of the transition matrix 
elements $I^{\cdot}_{\cdot}$, Eqs. (12) can be substantially
simplified. To do this, we define new  quantities that are 
convolutions of functions $C^{\cdot}_{\cdot}$ with overlap 
integral $I^{\cdot}_{\cdot}$ over only one variable, $K$ or $L$,
that  is
\begin{equation}
Z_L=\sum_{K_1}I^{2K_1}_{2K}C^{2L-1}_{2K_1}, \qquad
Z_K=\sum_{L_1}I^{2L}_{2L_1-1}C^{2L_1-1}_{2K}~, 
\end{equation}
\begin{displaymath}
Y_L=\sum_{K_1}I^{2K_1-1}_{2K}C^{2L-1}_{2K_1-1}~, \qquad
Y_K=\sum_{L_1}I^{2L-1}_{2L_1-1}C^{2L_1-1}_{2K-1}~, 
\end{displaymath}
\begin{displaymath}
X_L=\sum_{L_1}I^{2L}_{2L_1-1}C^{2L_1}_{2K}, \qquad
X_K=\sum_{L_1}I^{2L-1}_{2L_1}C^{2L_1}_{2K}~.
\end{displaymath}
Inserting Eqs. (14) into Eqs. (12), we obtain
\begin{equation}
C^{2L-1}_{2K}=\frac{1}{\mu H+y+x+\Delta}
\Bigl \{ I-Z_L+Y_L+X_K \Bigr \}~, 
\end{equation}
\begin{displaymath}
C^{2L-1}_{2K-1}=\frac{1}{y+x+\Delta}
\Bigl \{-I+Z_L-Y_L+Y_K \Bigr \}~, 
\end{displaymath}
\begin{displaymath}
C^{2L}_{2K}=\frac{1}{y+x+\Delta} Z_K~, 
\end{displaymath}
where $I$ is the value of the transition matrix element $I^{\cdot}_{\cdot}$
for states near the Fermi surface. Now from Eqs. (14) and (15) we can obtain 
a complete set of equations for the quantities $Z_{K,L};Y_{K,L};X_{K,L}$ only. 
In addition, the quantities $X_{K,L}$ are very simply related to 
$Z_{K,L};Y_{K,L}$. Eliminating them, we obtain a set equations for just 
the quantities $Z_{K,L};Y_{K,L}$:
\begin{equation}
Z_L \Bigl (
1+g\ln \frac{\varepsilon_F}{\mu H+y+\Delta} \Bigr )
-Y_Lg\ln \frac{\varepsilon_F}{\mu H+y+\Delta}=
\end{equation}
\begin{displaymath}
Ig\ln\frac{\varepsilon_F}{\mu H+y+\Delta}+g^2
\int\limits^{\varepsilon_F}_0
\frac{dxZ_K\ln\frac{A\varepsilon_F}{x+\Delta}}
{\mu H+y+x+\Delta}~,
\end{displaymath}
\begin{displaymath}
Z_K \Bigl (
1-g^2\ln \frac{A\varepsilon_F}{x+\Delta}
\ln\frac{A\varepsilon_F}{\mu H+x+\Delta} \Bigr )=
\end{displaymath}
\begin{displaymath}
Ig\ln\frac{A\varepsilon_F}{\mu H+x+\Delta}-g
\int\limits^{A\varepsilon_F}_0
\frac{dy(Z_L-Y_L)}
{\mu H+y+x+\Delta}~,
\end{displaymath}
\begin{displaymath}
Y_L \Bigl (
1+g\ln \frac{\varepsilon_F}{y+\Delta} \Bigr )-
Z_Lg\ln \frac{\varepsilon_F}{y+\Delta}=
-Ig\ln\frac{\varepsilon_F}{y+\Delta}+g
\int\limits^{\varepsilon_F}_0
\frac{dxY_K}
{y+x+\Delta}~,
\end{displaymath}
\begin{displaymath}
Y_K \Bigl (
1-g\ln \frac{A\varepsilon_F}{x+\Delta} \Bigr )=
-Ig\ln\frac{A\varepsilon_F}{x+\Delta}+g
\int\limits^{A\varepsilon_F}_0
\frac{dy(Z_L-Y_L)}
{y+x+\Delta}~.
\end{displaymath}

Equations (16) are valid for both signs of the 
interaction constant $g$. But its solutions are 
substantially different for $g<0$ and $g>0$. Consider 
first the case $g<0$ (attractive interaction in the Kondo problem). 
In such a case, the quantities $Z_L,~Y_L$  are large in comparison 
with $Z_K$ and $Y_K$. To obtain this, we introduce a formal definition 
of "Kondo" temperature $T_c$,
\begin{equation}
|g|\ln\frac{\varepsilon_F}{T_c}=1/2.
\end{equation}
Now we also put 
\begin{equation}
T_L(y)=Z_L-Y_L.
\end{equation}
Eliminating terms $Z_K,~Y_K$ from (16), we obtain one 
equation the quantity $T_L$:
\begin{equation}
T_{L}(y)=\frac{1}
{1+g\ln\frac{\varepsilon_F}{y+\Delta}+g\ln\frac{\varepsilon_F}
{\mu H+y+\Delta}}\cdot
\Biggl \{ \Biggr. Ig\Bigl (\ln\frac{\varepsilon_F}{y+\Delta}+
\ln\frac{\varepsilon_F}{\mu H+y+\Delta}\Bigr )
\end{equation}
\begin{displaymath}
+\frac{Ig}{2}\int\limits^{\varepsilon_F}_{0}dx
\Bigl ( \frac{1}{\mu H+y+x+\Delta}+
\frac{1}{y+x+\Delta} \Bigr )
\Biggl [ \Biggr. 
\frac{g^2\ln\frac{A\varepsilon_F}{x+\Delta}
\ln\frac{A\varepsilon_F}{\mu H+x+\Delta}}
{1-g^2\ln\frac{A\varepsilon_F}{x+\Delta}
\ln\frac{A\varepsilon_F}{\mu H+x+\Delta}}
\end{displaymath}
\begin{displaymath}
+\Biggl.  
\frac{g\ln\frac{A\varepsilon_F}{x+\Delta}}
{1-g\ln\frac{A\varepsilon_F}{x+\Delta}} \Biggr ]-
\frac{g^2}{2}\int\limits^{\varepsilon_F}_0dx
\Bigl ( \frac{1}{\mu H+y+x+\Delta}+
\frac{1}{y+x+\Delta} \Bigr ) 
\end{displaymath}
\begin{displaymath}
\times \Biggl [ \Biggr. 
\frac{g\ln\frac{A\varepsilon_F}{x+\Delta}}
{1-g^2\ln\frac{A\varepsilon_F}{x+\Delta}
\ln\frac{A\varepsilon_F}{\mu H+x+\Delta}} 
\end{displaymath}
\begin{displaymath}
\int\limits^{A\varepsilon_F}_0
\frac{dy_1T_L(y_1)}{\mu H+y_1+x+\Delta}+
\frac{1}{1-g\ln\frac{A\varepsilon_F}{x+\Delta}}
\int\limits^{A\varepsilon_F}_0
\frac{dy_1T_L(y_1)}{y_1+x+\Delta}
\Biggl. \Biggr ]
\Biggl. \Biggr \}
\end{displaymath}

It can be shown that the last term in Eq. (19) can be omitted, because 
it is small in the parameter $(g|\ln (1/|g|)$. We then obtain 
from Eqs. (17) and (19)
\begin{equation}
T_{L}(y)=
\frac{1}{|g|\ln\Bigl (
\frac{(y+\Delta)(\mu H+y+\Delta)}{T^2_c}\Bigr )}
\Bigl \{
-I-I\int\limits^{1/2}_0dt\Bigl (\frac{t^2}{1-t^2}-\frac{t}{1-t} \Bigr )
\Bigr \}.
\end{equation}
We finally obtain
\begin{equation}
T_L(y)=
\frac{-I\beta}
{|g|\ln \Bigl (
\frac{(y+\Delta)(\mu H+y+\Delta)}{T_c^2} \Bigr )}, \qquad
\beta=\frac{1}{2}\ln 3+\ln(3/2).
\end{equation}
Inserting Eqs. (18) and (21) into Eq. (15), we obtain expressions 
for coefficients $C^{2L-1}_{2K},~C^{2L-1}_{2K-1}$:
\begin{equation}
C^{2L-1}_{2K}=-
\frac{T_L(y)}{\mu H+y+x+\Delta}~, \qquad 
C^{2L-1}_{2K-1}=
\frac{T_L(y)}{y+x+\Delta}.
\end{equation}
Now we can determine the value of $\Delta$. Equations (10) and (11) 
should give the same value for average spin $\langle S_z\rangle$. This 
condition, with the help of Eqs. (4) and (22), gives
\begin{equation}
\beta\int\limits^{\infty}_0
\frac{dy}{(\mu H+y+\Delta)\ln^2
\Bigl ( \frac{(y+\Delta)(\mu H+y+\Delta)}{T^2_c} \Bigr )} \Biggr /
\Biggl (
1+\frac{\beta^2}{\ln \Bigl ( \frac{\Delta (\mu H+\Delta)}{T^2_c} \Bigr )}
\Biggr )=
\end{equation}
\begin{displaymath}
\frac{\partial\Delta}{\partial\mu H}\cdot
\frac{1}{\ln\frac{\Delta (\mu H+\Delta)}{T^2_c}}+
\int\limits^{\infty}_0
\frac{dy}{(\mu H+y+\Delta)\ln^2
\Bigl (\frac{(y+\Delta)(\mu H+y+\Delta)}{T^2_c} \Bigr )}.
\end{displaymath}
We seek a  solution of Eq. (23) in the form
\begin{equation}
\Delta (\mu H+\Delta)=T^2_c(1+\gamma), \qquad 0<\gamma\ll 1.
\end{equation}
Terms proportional to  $\gamma^{-1}$ cancel on the right-hand side 
of Eq. (23). This condition yields
\begin{equation}
\frac{\partial\Delta}{\partial\mu H}+
\frac{T^2_c}{(\mu H+\Delta)(\mu H+2\Delta)}=0.
\end{equation}
The solution of this equation is
\begin{equation}
\Delta (\mu H+\Delta )=T^2_c,
\end{equation}
$$
\quad\qquad\qquad\qquad \Delta =-\frac{\mu H}{2}+
\Biggl ( \Bigl ( \frac{\mu H}{2} \Bigr )^2+T^2_c \Biggr )^{1/2},
\quad\qquad\qquad\qquad\qquad\qquad (26a)
$$
and confirms our conjecture (24) about it. Of course, Eqs. (24) have two 
solutions for $\Delta$. One is given by Eq. (26a) (ground state),
and the other is
$$
\quad\qquad\qquad\qquad
\Delta = -\frac{\mu H}{2}-\Biggl ( \Bigl (
\frac{\mu H}{2} \Bigr )^2+T^2_c \Biggr )^{1/2}.
\quad\qquad\qquad\qquad\qquad\qquad  (26b)
$$
Solution (26b) for $\Delta$ corresponds to the excited state. 
In the limit $\mu H\gg T_c$, this state transforms to a state with spin 
orientation along the magnetic field. The excited state is separated from 
the ground state by a "gap" $2\Biggl ( \Bigl (
\frac{\mu H}{2} \Bigr )^2+T^2_c \Biggr )^{1/2}$. The gap results in the 
independence of the position of the maximum of impurity heat capacity 
from the magnetic field in the range $\mu H\ll T_c$ 
(Schottky anomaly). Such a residual Schottky anomaly is always present 
in experiments$^5$. In the Sec. 5 we will show 
that renormalization of the term $\mu H$ in (7) leads to a 
change from  $\mu H$ in Eq. (27) to $\mu\tilde H$ 
defined by Eq. (43). As a result, we obtain the mean spin 
$\langle S_z\rangle$ as an implicit  function of the magnetic field 
$\mu H$.

An attempt to obtain such an equation  at nonzero  
temperature was made in Ref. 7. But the mean field 
approximation used there is incorrect for the problem considered. 

In Appendix D we show that the right-hand side of (7)
leads to renormalization of the coefficients in Eq. 16, but does not
alter the main result of the paper, Eqs. (27) and (43). Of course,  
renormalization changes Eq. (17) for the Kondo temperature. 
The quantity 
$\gamma$ can be found only from correction terms to Eqs. (20) and (22). 
Fortunately, we do not need these  correction terms, because in the  
leading approximation, $\gamma$ also drops out of Eq. (11) for the spin 
value. With the help of Eqs. (11), (22), and (24), we obtain
\begin{displaymath}
\langle S_z\rangle=\frac{\mu H}{2}\cdot
\int\limits^{\infty}_0
\frac{dy}{(y+\Delta)(y+\mu H+\Delta)(\gamma +y(\mu H+2\Delta)/T^2_c)^2}
\Biggr / 1/\gamma
\end{displaymath}
\begin{equation}
=\frac{\mu H}{4 \Bigl ( T^2_c+(\mu H/2)^2\Bigr  )^{1/2}}.
\end{equation}
Equation (27) is in good agreement with the experimental data of Ref. 4.

\vspace{0.5cm}
\section{Ferromagnetic case ${\bf (g>0)}$}

As mentioned  above,  Eqs. (16) are valid for both signs of 
the "interaction" constant $g$. In the case $g>0$, we can define the 
characteristic energy of the problem to be the Kondo temperature $T_c$ by the 
relation
\begin{equation}
g\ln \frac{A\varepsilon_F}{T_c}=1.
\end{equation}
For $g>0$, the quantities $Z_K,~Y_K,~X_K$ are large in comparison 
with $Z_L,~Y_L,~X_L$. We can eliminate $Z_L,~Y_L$ 
from Eqs. (16). As a result, we have 
\begin{equation}
Z_K(1-g^2\ln\frac{A\varepsilon_F}{x+\Delta}
\ln\frac{A\varepsilon_F}{\mu H+x+\Delta}=
Ig\ln\frac{A\varepsilon_F}{\mu H+x+\Delta}
\end{equation}
\begin{displaymath}
-g\int\limits^{A\varepsilon_F}_0
\frac{dy}{\mu H+y+x+\Delta} \cdot
\frac{1}{1+g\ln\frac{\varepsilon_F}{y+\Delta}+
g\ln\frac{\varepsilon_F}{\mu H+y+\Delta}} \cdot
\Biggl [ \Biggr. Ig\ln \frac{\varepsilon^2_F}{(y+\Delta)(\mu H+y+\Delta)}
\end{displaymath}
\begin{displaymath}
+g^2\int\limits^{\varepsilon_F}_0
\frac{dx_1Z_K(x_1)\ln\frac{A\varepsilon_F}{x_1+\Delta}} 
{\mu H+y+x_1+\Delta}-g\int\limits^{\varepsilon_F}_0
\frac{dx_1Y_K(x_1)}{y+x_1+\Delta} \Biggl. \Biggr ]~,
\end{displaymath}
\begin{displaymath}
Y_K(1-g\ln\frac{A\varepsilon_F}{x+\Delta})=
-Ig\ln\frac{A\varepsilon_F}{x+\Delta}
\end{displaymath}
\begin{displaymath}
+g\int\limits^{A\varepsilon_F}_0
\frac{dy}{y+x+\Delta} \cdot
\frac{1}{1+g\ln\frac{\varepsilon_F}{y+\Delta}+
\ln\frac{\varepsilon_F}{\mu H+y+\Delta}} \cdot
\Biggl [ \Biggr. Ig\ln \frac{\varepsilon^2_F}{(y+\Delta)(\mu H+y+\Delta)}
\end{displaymath}
\begin{displaymath}
+g^2\int\limits^{\varepsilon_F}_0
\frac{dx_1Z_K(x_1)\ln\frac{A\varepsilon_F}{x_1+\Delta}}
{\mu H+y+x_1+\Delta}-g\int\limits^{\varepsilon_F}_0
\frac{dx_1Y_K(x_1)}{y+x_1+\Delta} \Biggl. \Biggr ]~.
\end{displaymath}

In the range $x\ll \varepsilon_F$, Eqs. (29) yield
the following values for  $Y_K,~Z_K$:
\begin{equation}
Y_K=-\frac{ID}{g\ln \bigl (
\frac{x+\Delta}{T_c} \bigr )}~, \qquad
Z_K=\frac{ID}{g\ln \bigl ( 
\frac{(x+\Delta)(\mu H+x+\Delta)}
{T^2_c} \bigr )}~,
\end{equation}
where $D$ is a number of order  unity. Inserting Eq. (30) into Eq. (15),
we obtain
\begin{equation}
C^{2L-1}_{2K}=\frac{1}{\mu H+y+x+\Delta}\cdot
\frac{ID}{g\ln \bigl ( \frac{(x+\Delta)(\mu H+x+\Delta)} 
{T^2_c} \bigr )}~,
\end{equation}
\begin{displaymath}
C^{2L-1}_{2K-1}=-\frac{1}{y+x+\Delta}\cdot
\frac{ID}{g\ln \bigl ( \frac{x+\Delta} 
{T_c} \bigr )}~,
\end{displaymath}
\begin{displaymath}
C^{2L}_{2K}=\frac{1}{y+x+\Delta}\cdot
\frac{ID}{g\ln \bigl ( \frac{(x+\Delta)(\mu H+x+\Delta)} 
{T^2_c} \bigr )}~.
\end{displaymath}

In the same way as in the case $g<0$,  with the help 
of Eqs. (10), (11), and (31), we obtain 
\begin{equation}
\frac{\partial\Delta}{\partial\mu H}
\Biggl [
\frac{1}{\ln\frac{\Delta}{T_c}}+
\frac{1}{\ln\frac{\Delta (\mu H+\Delta)}{T^2_c}} \Biggr ]=
\end{equation}
\begin{displaymath}
-\Biggl [ 1-\frac{D^2}
{1+D^2 \Bigl ( \frac{1}{\ln\Delta\bigl / T_c}+
\frac{1}{\ln \frac{\Delta (\mu H+\Delta)}{T^2_c}} \Bigr )} \Biggr ]
\int\limits^{\infty}_0
\frac{dx}{(x+\Delta+\mu H)\ln^2 \Bigl (
\frac{(x+\Delta)(\mu H+x+\Delta)}{T^2_c} \Bigr )}.
\end{displaymath}
The solution of this equation is 
\begin{equation}
\Delta\equiv T_c.
\end{equation}
Equation (33) means that in the leading approximation, the 
spin value in the magnetic field is saturated,
\begin{equation}
\langle S_z\rangle =\frac{1}{2}.
\end{equation}

Correction terms to Eq. (34) come only from an energy range 
$\varepsilon$ of order  $\varepsilon\sim\varepsilon_F\exp (-1/g^2)$.
Note that a similar energy scale also arises  in the problem
considered by Nozieres and Dominicis$^6$. Our conjecture 
is that in temperature range
\begin{equation}
T^2_c/\varepsilon_F~\ll T\ll T_c,
\end{equation}
the leading  correction to the average spin arises from the 
cutoff of integrals with respect to  energy in expression (11) over an 
energy range of order $T$. If such an assumption is true, then the 
average spin
in the magnetic field   $\mu H \gg T$ is 
\begin{equation}
\langle S_z\rangle=\frac{1}{2}-\frac{T}{T_c}
\int\limits^{\infty}_0
\frac{dx}
{(x+1+\mu H/T_c)\ln^2\Bigl ( (1+x)(x+1+\mu H/T_c) \Bigr )}
\end{equation}
\begin{displaymath}
=\frac{1}{2}-\frac{T}{4T_c}
\int\limits^{\infty}_{\ln(1+\mu H/T_c)}
\frac{dz}
{\Bigl [ z+1/2\ln(1-(\mu H/T_c)e^{-z})\Bigr ]^2}.
\end{displaymath}
In the limiting cases of weak $(\mu H \ll T_c)$ and strong 
$(\mu H \gg T_c)$ magnetic fields, the average  spin is 
\begin{equation}
\langle S_z\rangle =\frac{1}{2} \Bigl ( 1-\frac{T}{\mu H} \Bigr ), \qquad
T\ll \mu H\ll T_c,
\end{equation}
\begin{displaymath}
\langle S_z\rangle =\frac{1}{2} \Bigl ( 1-\frac{T}
{2T_c\ln \bigl ( \frac{\mu H}{T_c} \bigr )} \Bigr ), \qquad
\mu H\gg T_c.
\end{displaymath}

\vspace{0.5cm}
\section{Self-energy terms 
${\bf \Sigma^{(1)}_{(K,L)},~\Sigma_{(K,L)}}$
in perturbation theory}

As mentioned in the Sec. 2,  
there are two self-energy terms in the problem under consideration,
$\Sigma^{(1)}_{(K,L)}$ and $\Sigma_{(K,L)}$.
In  second-order perturbation theory, they coincide. They start to 
differ in third-order in the coupling constant. In third-order 
perturbation theory, we otain from Appendix A
\begin{equation}
\Sigma^{(1)}_{(K,L)}-\Sigma_{(K,L)}=
I^{2K_1}_{2L_1}I^{2L_1}_{2L_2}I^{2L_2}_{2K_1}
\end{equation}
\begin{displaymath}
\times\Biggl (
\frac{1}
{(\mu H+\varepsilon_L+\varepsilon_{L_1}-\varepsilon_K-\varepsilon_{K_1})
(\mu H+\varepsilon_L+\varepsilon_{L_2}-\varepsilon_K-\varepsilon_{K_1})}
\end{displaymath}
\begin{displaymath}
-\frac{1}
{(\varepsilon_L+\varepsilon_{L_1}-\varepsilon_K-\varepsilon_{K_1})
(\varepsilon_L+\varepsilon_{L_2}-\varepsilon_K-\varepsilon_{K_1})}
\Biggr ).
\end{displaymath}

A simple calculation of sums in Eq. (38) leads to 
\begin{equation}
\Bigl ( \Sigma^{(1)}_{(K,L)}-\Sigma_{(K,L)}\Bigr )_
{\varepsilon_L=\varepsilon_K=\varepsilon_F}=
-\mu Hg^3\ln^2\Bigl (\frac{\varepsilon_F}{\varepsilon_c} \Bigr ),
\end{equation}
where $\varepsilon_c$ is the cutoff energy. In  Appendix C, we obtain 
the following term in expansion (39) for the self-energy:
\begin{equation}
\Bigl (\Sigma^{(1)}_{(K,L)}-\Sigma_{(K,L)}\Bigr )
_{\varepsilon_K=\varepsilon_L=\varepsilon_F}=
-\mu Hg^3\ln^2
\Bigl (\frac{\varepsilon_F}{\varepsilon_c} \Bigr ) +
2\mu Hg^4\ln^3 \Bigl (\frac{\varepsilon_F}{\varepsilon_c} \Bigr )-...
\end{equation}
Comparison with the expression for $\delta E$ obtained in perturbation 
theory shows that 
\begin{equation}
\delta\Sigma=
\Bigl ( \Sigma^{(1)}_{(K,L)}-\Sigma_{(K,L)}\Bigr )_
{\varepsilon_K=\varepsilon_L=\varepsilon_F}
\end{equation}
\begin{displaymath}
=\mu Hg\ln \Bigl (\frac{\varepsilon_F}{\varepsilon_c}\Bigr )
\Biggl [ -\frac{\partial\delta E}{\partial\mu H} \Biggr ]
\end{displaymath}
\begin{displaymath}
=-\frac{\mu H}{2} \Bigl ( -\frac{1}{2}+\langle S_z \rangle \Bigr )~.
\end{displaymath}
To obtain Eq. (41) we used Eqs. (17), (10), and an assumption that 
$\varepsilon_c\sim T_c$.

Equation (41) means that some corrections should be made in the first of
Eqs. (12). Specifically, $\mu H$ in the first Eqs. (12) should be
corrected by $\delta\Sigma$:
\begin{equation}
\mu H \to \mu H-\delta\Sigma =\mu \tilde H
\end{equation}
The main result of this correction is a decrease in the initial 
slope of the magnetic field dependence of the average spin value 
by 3/4. This phenomenon was probably found in the 
experimental Ref. 4 (Figs 8 and 9). The average spin 
$\langle S_z\rangle$ is given by Eq. (27) with the substitution
\begin{equation}
\mu H \to \mu\tilde H=\mu H -
\frac{\mu H}{2} \Bigl ( \frac{1}{2}-\langle S_z\rangle \Bigr ).
\end{equation}
This equation determines $\langle S_z\rangle$ as an implicit
function of $\mu H$. From Eqs. (27) and (43), 
we find that $\langle S_z\rangle$ as a function of $\mu H$ has an 
a inflection point at $\mu H/2T_c=0.2426$. Such an 
inflection point was obtained in Ref. 4.

\vspace{0.5cm}
\section{Conclusion}

Thus, we show that at zero temperature and finite 
magnetic field $\mu H \ll \varepsilon_F$, a singularity esists in the  
convolution of amplitudes $C^{2L-L}_{2K_1}$ and $C^{2L-1}_{2K_1-1}$ 
over energy $\varepsilon_{K_1}$ with amplitude $I^{2K_1}_{2K}$. As a
result, in the high magnetic field region $\mu H\gg T_c$ the correction 
to the spin impurity value is proportional to $(T_c/\mu H)^2$
instead of $1/\ln (\mu H/T_c)$, as predicted in Refs. 1-3. 
We also find that renormalization of the 
magnetic field discussed in Sec. 5 leads to  
an inflection point in the dependence of spin impurity on the magnetic 
field. The initial slope is a function of $z$, which  
enters into the definition of the Kondo temperature (see Appendix D). Our 
consideration shows that the interaction of the spin of 
an impurity with at  
electron gas does not lead to the appearance of the 
localized state, as  
assumed in Refs. 8-10. The Kondo temperature $T_c$ 
is given by Eq. (D.7), where $z$ is the root of the equation 
\begin{equation}
f(z)=0.
\end{equation}
We find here three terms in the expansion of $f$
in Taylor series (Eq. (D.8)). This equation was also studied in 
Refs. 8 and 11. Our result for the first two terms in Eq. (44) 
coincide with the result of Ref. 11, because this is also the result of 
parquet approximation. But, our consideration (Eq. 44) 
is conceptually closer to the Ref. 8. The difference even in the 
second term is probably  related to the assumption of Ref. 8
that in the problem under consideration there is a localized state 
with spin 1/2. 

In fact, such a localized state does not exist. Without  
interaction there are two states associted with impurity spin 1/2. 
In zero magnetic field, these two states are degenerate. Interaction 
removes such a degeneracy and the splitting energy is $2T_c$. 
Of course, interaction does not change the number of states, as  
in our consideration, and is not fulfilled in Ref. 8. 
Note also that the driving term is Refs. 8-10 missing. 

Nevertheless, the average value of spin of impurity $\langle S_z\rangle$
as a function of magnetic field found in Refs. 9 and 10 
coincides with our result except for the effect of renormalization of the 
magnetic field (Sec. 5) and the expression for the Kondo temperature. 

\vspace{0.2cm}
The authors thank Prof. P. Fulde and Prof. A.I. Larkin for  
helpful discussions. We thank Prof. P. Fulde for hospitality 
at the Max-Planck-Institute for Complex Systems (Dresden). 
The research of Yu.N.O. was supported by  CRDF Grant RP1-194. 
The research of A.M.D. is supported by INTAS and the Russian 
Foumdation for Basic Research (Grant 95-553). 

\newpage

\def\theequation{\thesection.\arabic{equation}}
\appendix
\setcounter{equation}{0}
\section{Appendix}

\vspace{0.5cm}
The wave function of a system consisting of one localized electron plus 
degenerate electron gas can be taken in the form
\begin{displaymath}
|\psi\rangle =|10;11;11...\rangle +
\sum C^{2L-1}_{2K}|01;\stackrel{2K}{10};\stackrel{2L-1}{10}\rangle
\end{displaymath}
\begin{displaymath}
+\sum C^{2L-1}_{2K-1}|10;\stackrel{2K-1}{01};\stackrel{2L-1}{10}\rangle+
\sum C^{2L}_{2K}|10;\stackrel{2K}{10};\stackrel{2L}{01}\rangle
\end{displaymath}
\begin{displaymath}
+\sum_{K_1<K} C^{2L_1;2L-1}_{2K_1;2K}
\hat N|01;\stackrel{2K_1}{10};\stackrel{2K}{10};
\stackrel{2L_1}{01};\stackrel{2L-1}{10}\rangle+
\sum C^{2L_1;2L-1}_{2K_1-1;2K}
\hat N|10;\stackrel{2K_1-1}{01};\stackrel{2K}{10};
\stackrel{2L_1}{01};\stackrel{2L-1}{10}\rangle
\end{displaymath}
\begin{displaymath}
+\sum_{L_1<L} C^{2L_1-1;2L-1}_{2K_1;2K-1}
\hat N|01;\stackrel{2K_1}{10};\stackrel{2K-1}{01};
\stackrel{2L_1-1}{10};\stackrel{2L-1}{10}\rangle
\end{displaymath}
\begin{displaymath}
+\sum_{K_1<K;L_1<L} C^{2L_1-1;2L-1}_{2K_1-1;2K-1}
\hat N|10;\stackrel{2K_1-1}{01};\stackrel{2K-1}{01};
\stackrel{2L_1-1}{10};\stackrel{2L-1}{10}\rangle
\end{displaymath}
\begin{displaymath}
+\sum_{K_1<K;L_1<L} C^{2L_1;2L}_{2K_1;2K}
|10;\stackrel{2K_1}{10};\stackrel{2K}{10};
\stackrel{2L_1}{01};\stackrel{2L}{01}\rangle
\end{displaymath}
\begin{displaymath}
+\sum_{K_2<K_1<K;L_2<L_1} C^{2L_1;2L_1;2L-1}_{2K_2;2K_1;2K}
\hat N|01;\stackrel{2K_2}{10};\stackrel{2K_1}{10};
\stackrel{2K}{10};\stackrel{2L_2}{01}\stackrel{2L_1}{01};
\stackrel{2L-1}{10}\rangle
\end{displaymath}
\begin{displaymath}
+\sum_{K_1<K;L_2<L_1} C^{2L_2;2L_1;2L-1}_{2K_2-1;2K_1;2K}
\hat N|10;\stackrel{2K_2-1}{01};\stackrel{2K_1}{10};
\stackrel{2K}{10};\stackrel{2L_2}{01}\stackrel{2L_1}{01};
\stackrel{2L-1}{10}\rangle
\end{displaymath}
\begin{displaymath}
+\sum_{K_2<K;L_1<L} C^{2L_2;2L_1-1;2L-1}_{2K_2-1;2K_1;2K-1}
\hat N|10;\stackrel{2K_2-1}{01};\stackrel{2K_1}{10};
\stackrel{2K-1}{01};\stackrel{2L_2}{01}\stackrel{2L_1-1}{10};
\stackrel{2L-1}{10}\rangle
\end{displaymath}
\begin{displaymath}
+\sum_{K_2<K_1;L_1<L} C^{2L_2;2L_1-1;2L-1}_{2K_2;2K_1;2K-1}
\hat N|01;\stackrel{2K_2}{10};\stackrel{2K_1}{10};
\stackrel{2K-1}{01};\stackrel{2L_2}{01}\stackrel{2L_1-1}{10};
\stackrel{2L-1}{10}\rangle
\end{displaymath}
\begin{displaymath}
+\sum_{K_1<K;L_2<L_1<L} C^{2L_2-1;2L_1-1;2L-1}_{2K_2;2K_1-1;2K-1}
\hat N|01;\stackrel{2K_2}{10};\stackrel{2K_1-1}{01};
\stackrel{2K-1}{01};\stackrel{2L_2-1}{10}\stackrel{2L_1-1}{10};
\stackrel{2L-1}{10}\rangle
\end{displaymath}
\begin{displaymath}
+\sum_{K_2<K_1<K;L_2<L_1<L} C^{2L_2-1;2L_1-1;2L-1}_{2K_2-1;2K_1-1;2K-1}
\hat N|10;\stackrel{2K_2-1}{01};\stackrel{2K_1-1}{01};
\stackrel{2K-1}{01};\stackrel{2L_2-1}{10}\stackrel{2L_1-1}{10};
\stackrel{2L-1}{10}\rangle
\end{displaymath}
\begin{equation}
+\sum_{K_2<K_1<K;L_2<L_1<L} C^{2L_2;2L_1;2L}_{2K_2;2K_1;2K}
\hat N|10;\stackrel{2K_2-1}{10};\stackrel{2K_1-1}{10};
\stackrel{2K-1}{10};\stackrel{2L_2}{01}\stackrel{2L_1}{01};
\stackrel{2L}{01}\rangle+ ...
\end{equation}
The notation here is the same as in the text. As we note above, there are 
$(2P+1)$ different symbols $C^{\cdot\cdot\cdot}_{\cdot\cdot\cdot}$
of order $P$. Inserting Eq. (A.1) into Eq. (3) for the wave function,
some simple but tedions calculations yield a set of equations for 
the 
coefficients $C^{\cdot\cdot}_{\cdot\cdot}$. The five equations for the 
$C^{\cdot\cdot}_{\cdot\cdot}$ are 
\begin{displaymath}
C^{2L-1}_{2K}I^{2L_1}_{2K_1}-C^{2L-1}_{2K_1}I^{2L_1}_{2K}-
C^{2L_1}_{2K}I^{2L-1}_{2K_1}+C^{2L_1}_{2K_1}I^{2L-1}_{2K}
\end{displaymath}
\begin{displaymath}
+(\mu H+\varepsilon_L+\varepsilon_{L_1}-\varepsilon_K-\varepsilon_{K_1}-
\delta E)C^{2L_1;2L-1}_{2K_1;2K}
\end{displaymath}
\begin{displaymath}
+\Biggl (
\sum_{K_2<K}C^{2L_1;2L-1}_{2K_2;2K}I^{2K_2}_{2K_1}-
\sum_{K<K_2}C^{2L_1;2L-1}_{2K;2K_2}I^{2K_2}_{2K_1} \Biggr )
\end{displaymath}
\begin{displaymath}
+\Biggl (
\sum_{K_1<K_2}C^{2L_1;2L-1}_{2K_1;2K_2}I^{2K_2}_{2K}-
\sum_{K_2<K_1}C^{2L_1;2L-1}_{2K_2;2K_1}I^{2K_2}_{2K} \Biggr )
\end{displaymath}
\begin{displaymath}
-\sum C^{2L_2;2L-1}_{2K_1;2K}I^{2L_1}_{2L_2}+
\Biggl (
\sum_{L_2<L_1}C^{2L_2;2L_1}_{2K_1;2K}I^{2L-1}_{2L_2}- 
\sum_{L_1<L_2}C^{2L_1;2L_2}_{2K_1;2K}I^{2L-1}_{2L_2}
\Biggr )
\end{displaymath}
\begin{displaymath}
-\Biggl (
\sum C^{2L_1;2L-1}_{2K_2-1;2K}I^{2K_2-1}_{2K_1}-
\sum C^{2L_1;2L-1}_{2K_2-1;2K_1}I^{2K_2-1}_{2K} \Biggr )
\end{displaymath}
\begin{displaymath}
-\sum_{K_1<K<K_2;L_2<L_1} C^{2L_2;2L_1;2L-1}_{2K_1;2K;2K_2}I^{2K_2}_{2L_2}
\end{displaymath}
\begin{displaymath}
+\sum_{K_1<K_2<K;L_2<L_1} C^{2L_2;2L_1;2L-1}_{2K_1;2K_2;2K}I^{2K_2}_{2L_2}-
\sum_{K_2<K_1;L_2<L_1} C^{2L_2;2L_1;2L-1}_{2K_2;2K_1;2K}I^{2K_2}_{2L_2}
\end{displaymath}
\begin{displaymath}
+\sum_{K_1<K<K_2;L_1<L_2} C^{2L_1;2L_2;2L-1}_{2K_1;2K;2K_2}I^{2K_2}_{2L_2}
\end{displaymath}
\begin{displaymath}
-\sum_{K_1<K_2<K;L_1<L_2} C^{2L_1;2L_2;2L-1}_{2K_1;2K_2;2K}I^{2K_2}_{2L_2}+
\sum_{K_2<K_1<K;L_1<L_2} C^{2L_1;2L_2;2L-1}_{2K_2;2K_1;2K}I^{2K_2}_{2L_2}
\end{displaymath}
\begin{displaymath}
+\sum_{L_2<L_1;K_1<K} C^{2L_2;2L_1;2L-1}_{2K_2-1;2K_1;2K}I^{2K_2-1}_{2L_2}-
\sum_{L_1<L_2;K_1<K} C^{2L_1;2L_2;2L-1}_{2K_2-1;2K_1;2K}I^{2K_2-1}_{2L_2}=0~,
\end{displaymath}
\begin{displaymath}
-I^{2L_1}_{2K_1-1}C^{2L-1}_{2K}+
C^{2L_1}_{2K}I^{2L-1}_{2K_1-1}-
\sum_{K_2<K}C^{2L_1;2L-1}_{2K_2;2K}I^{2K_2}_{2K_1-1}+
\sum_{K<K_2}C^{2L_1;2L-1}_{2K;2K_2}I^{2K_2}_{2K_1-1}
\end{displaymath}
\begin{displaymath}
+(\varepsilon_L+\varepsilon_{L_1}-\varepsilon_K-\varepsilon_{K_1}-\delta E)
C^{2L_1;2L-1}_{2K_1-1;2K}+
\sum C^{2L_1;2L-1}_{2K_2-1;2K}I^{2K_2-1}_{2K_1-1}-
\sum C^{2L_1;2L_2-1}_{2K_1-1;2K}I^{2L-1}_{2L_2-1}
\end{displaymath}
\begin{displaymath}
+\sum_{L_2<L} C^{2L_2-1;2L-1}_{2K;2K_1-1}I^{2L_1}_{2L_2-1}-
\sum_{L<L_2} C^{2L-1;2L_2-1}_{2K;2K_1-1}I^{2L_1}_{2L_2-1}
\end{displaymath}
\begin{displaymath}
+\sum_{L_2<L;K<K_2} C^{2L_1;2L_2-1;2L-1}_{2K;2K_2;2K_1-1}I^{2K_2}_{2L_2-1}-
\sum_{L_2<L;K_2<K} C^{2L_1;2L_2-1;2L-1}_{2K_2;2K;2K_1-1}I^{2K_2}_{2L_2-1}
\end{displaymath}
\begin{displaymath}
-\sum C^{2L_1;2L-1;2L_2-1}_{2K;2K_2;2K_1-1}I^{2K_2}_{2L_2-1}+
\sum_{L<L_2;K_2<K} C^{2L_1;2L-1;2L_2-1}_{2K_2;2K;2K_1-1}I^{2K_2}_{2L_2-1}
\end{displaymath}
\begin{displaymath}
-\sum_{L_2<L;K_1<K_2} C^{2L_1;2L_2-1;2L-1}_{2K_1-1;2K;2K_2-1}I^{2K_2-1}_{2L_2-1}+
\sum_{L_2<L;K_2<K_1} C^{2L_1;2L_2-1;2L-1}_{2K_2-1;2K;2K_1-1}I^{2K_2-1}_{2L_2-1}
\end{displaymath}
\begin{equation}
+\sum_{L<L_2;K_1<K_2} C^{2L_1;2L-1;2L_2-1}_{2K_1-1;2K;2K_2-1}I^{2K_2-1}_{2L_2-1}-
\sum_{L<L_2;K_2<K_1} C^{2L_1;2L-1;2L_2-1}_{2K_2-1;2K;2K_1-1}I^{2K_2-1}_{2L_2-1}
=0~,
\end{equation}
\begin{displaymath}
C^{2L-1}_{2K-1}I^{2L_1-1}_{2K_1}-C^{2L_1-1}_{2K-1}I^{2L-1}_{2K_1}+
\sum C^{2L_2;2L-1}_{2K-1;2K_1}I^{2L_1-1}_{2L_2}-
\sum C^{2L_2;2L_1-1}_{2K-1;2K_1}I^{2L-1}_{2L_2}
\end{displaymath}
\begin{displaymath}
+(\mu H+\varepsilon_L+\varepsilon_{L_1}-\varepsilon_K-\varepsilon_{K_1}-
\delta E)C^{2L_1-1;2L-1}_{2K_1;2K-1}+
\sum C^{2L_1-1;2L-1}_{2K_2;2K-1}I^{2K_2}_{2K_1}
\end{displaymath}
\begin{displaymath}
+\sum_{K<K_2}C^{2L_1-1;2L-1}_{2K-1;2K_2-1}I^{2K_2-1}_{2K_1}-
\sum_{K_2<K}C^{2L_1-1;2L-1}_{2K_2-1;2K-1}I^{2K_2-1}_{2K_1}+
\sum_{K_1<K_2} C^{2L_2;2L_1-1;2L-1}_{2K_1;2K_2;2K-1}I^{2K_2}_{2L_2}
\end{displaymath}
\begin{displaymath}
-\sum_{K_2<K_1}C^{2L_2;2L_1-1;2L-1}_{2K_2;2K_1;2K-1}I^{2K_2}_{2L_2}-
\sum_{K<K_2}C^{2L_2;2L_1-1;2L-1}_{2K-1;2K_1;2K_2-1}I^{2K_2-1}_{2L_2}
\end{displaymath}
\begin{displaymath}
+\sum_{K_2<K} C^{2L_2;2L_1-1;2L-1}_{2K_2-1;2K_1;2K-1}I^{2K_2-1}_{2L_2}=0~,
\end{displaymath}
\begin{displaymath}
(\varepsilon_L+\varepsilon_{L_1}-\varepsilon_K-\varepsilon_{K_1}-\delta E)
C^{2L_1;2L}_{2K_1;2K}+
\sum C^{2L;2L_2-1}_{2K_1;2K}I^{2L_1}_{2L_2-1}-
\sum C^{2L_1;2L_2-1}_{2K_1;2K}I^{2L}_{2L_2-1}
\end{displaymath}
\begin{displaymath}
-\sum_{K_1<K<K_2} C^{2L_1;2L;2L_2-1}_{2K_1;2K;2K_2}I^{2K_2}_{2L_2-1}+
\sum_{K_1<K_2<K} C^{2L_1;2L;2L_2-1}_{2K_1;2K_2;2K}I^{2K_2}_{2L_2-1}
\end{displaymath}
\begin{displaymath}
-\sum_{K_2<K_1<K} C^{2L_1;2L;2L_2-1}_{2K_2;2K_1;2K}I^{2K_2}_{2L_2-1}+
\sum C^{2L_1;2L;2L_2-1}_{2K_2-1;2K_1;2K}I^{2K_2-1}_{2L_2-1}=0 ~,
\end{displaymath}
\begin{displaymath}
-I^{2L_1-1}_{2K_1-1}C^{2L-1}_{2K-1}+
C^{2L-1}_{2K_1-1}I^{2L_1-1}_{2K-1}+
C^{2L_1-1}_{2K-1}I^{2L-1}_{2K_1-1}-
C^{2L_1-1}_{2K_1-1}I^{2L-1}_{2K-1}
\end{displaymath}
\begin{displaymath}
+(\varepsilon_L+\varepsilon_{L_1}-\varepsilon_K-\varepsilon_{K_1}-\delta E)
C^{2L_1-1;2L-1}_{2K_1-1;2K-1}
\end{displaymath}
\begin{displaymath}
-\Biggl (
\sum C^{2L_1-1;2L-1}_{2K_2;2K-1}I^{2K_2}_{2K_1-1}-
\sum C^{2L_1-1;2L-1}_{2K_2;2K_1-1}I^{2K_2}_{2K-1} \Biggr )
\end{displaymath}
\begin{displaymath}
+\Biggl (
\sum_{K_2<K} C^{2L_1-1;2L-1}_{2K_2-1;2K-1}I^{2K_2-1}_{2K_1-1}-
\sum_{K_2<K_1} C^{2L_1-1;2L-1}_{2K_2-1;2K_1-1}I^{2K_2-1}_{2K-1} \Biggr )
\end{displaymath}
\begin{displaymath}
-\Biggl (
\sum_{K<K_2} C^{2L_1-1;2L-1}_{2K-1;2K_2-1}I^{2K_2-1}_{2K_1-1}-
\sum_{K_1<K_2} C^{2L_1-1;2L-1}_{2K_1-1;2K_2-1}I^{2K_2-1}_{2K-1} \Biggr )
\end{displaymath}
\begin{displaymath}
-\Biggl (
\sum_{L_2<L} C^{2L_2-1;2L-1}_{2K_1-1;2K-1}I^{2L_1-1}_{2L_2-1}-
\sum_{L_2<L_1} C^{2L_2-1;2L_1-1}_{2K_1-1;2K-1}I^{2L-1}_{2L_2-1} \Biggr )
\end{displaymath}
\begin{displaymath}
+\Biggl (
\sum_{L<L_2} C^{2L-1;2L_2-1}_{2K_1-1;2K-1}I^{2L_1-1}_{2L_2-1}-
\sum_{L_1<L_2} C^{2L_1-1;2L_2-1}_{2K_1-1;2K-1}I^{2L-1}_{2L_2-1} \Biggr )
\end{displaymath}
\begin{displaymath}
-\sum_{L_2<L_1<L} C^{2L_2-1;2L_1-1;2L-1}_{2K_2;2K_1-1;2K-1}I^{2K_2}_{2L_2-1}+
\sum_{L_1<L_2<L} C^{2L_1-1;2L_2-1;2L-1}_{2K_2;2K_1-1;2K-1}I^{2K_2}_{2L_2-1}
\end{displaymath}
\begin{displaymath}
-\sum_{L_1<L<L_2} C^{2L_1-1;2L-1;2L_2-1}_{2K_2;2K_1-1;2K-1}
I^{2K_2}_{2L_2-1}+
\sum_{L_2<L_1<L;K<K_2} C^{2L_2-1;2L_1-1;2L-1}_{2K_1-1;2K-1;2K_2-1}
I^{2K_2-1}_{2L_2-1}
\end{displaymath}
\begin{displaymath}
-\sum_{K_1<K_2<K;L_2<L_1} C^{2L_2-1;2L_1-1;2L-1}_{2K_1-1;2K_2-1;2K-1}
I^{2K_2-1}_{2L_2-1}+
\sum_{K_2<K_1;L_2<L_1} C^{2L_2-1;2L_1-1;2L-1}_{2K_2-1;2K_1-1;2K-1}
I^{2K_2-1}_{2L_2-1}
\end{displaymath}
\begin{displaymath}
-\sum_{K<K_2;L_1<L_2<L} C^{2L_1-1;2L_2-1;2L-1}_{2K_1-1;2K-1;2K_2-1}
I^{2K_2-1}_{2L_2-1}+
\sum_{K_1<K_2<K;L_1<L_2<L} C^{2L_1-1;2L_2-1;2L-1}_{2K_1-1;2K_2-1;2K-1}
I^{2K_2-1}_{2L_2-1}
\end{displaymath}
\begin{displaymath}
-\sum_{K_2<K_1;L_1<L_2<L} C^{2L_1-1;2L_2-1;2L-1}_{2K_2-1;2K_1-1;2K-1}
I^{2K_2-1}_{2L_2-1}+
\sum_{K_1<K<K_2;L<L_2} C^{2L_1-1;2L-1;2L_2-1}_{2K_1-1;2K-1;2K_2-1}
I^{2K_2-1}_{2L_2-1}
\end{displaymath}
\begin{displaymath}
-\sum_{L<L_2;K_1<K_2<K} C^{2L_1-1;2L-1;2L_2-1}_{2K_1-1;2K_2-1;2K-1}
I^{2K_2-1}_{2L_2-1}+
\sum_{K_2<K_1;L<L_2} C^{2L_1-1;2L-1;2L_2-1}_{2K_2-1;2K_1-1;2K-1}
I^{2K_2-1}_{2L_2-1}=0
\end{displaymath}

Equations (A.2) are exact. 

\newpage
\def\theequation{\thesection.\arabic{equation}}
\setcounter{equation} {0}
\section{Appendix}

Our purpose is to obtain an expression for the self-energy terms
$\Sigma^{(1)}_{(K,L)}$ and $\Sigma_{(K,L)}$ in fourth-order  
perturbation theory. To do this we should obtain equations on the 
quantities $C^{\cdot\cdot\cdot}_{\cdot\cdot\cdot}$ in the "leading"
approximation. That is, we can omit in such a system of equations the 
terms corresponding to "scattering" of terms 
$C^{\cdot\cdot\cdot}_{\cdot\cdot\cdot}$ and connection terms with 
quantities $C^{\cdot\cdot\cdot\cdot}_{\cdot\cdot\cdot\cdot}$. 
Really, we need only six equations in the six quantities entering
into Eqs. (A.2). The required system can be obtained 
from Eqs. (3) and (A.1). These equations are 
\begin{displaymath}
(\mu H+\varepsilon_L+\varepsilon_{L_1}+\varepsilon_{L_2}-
\varepsilon_K-\varepsilon_{K_1}-\varepsilon_{K_2})
C^{2L_2;2L_1;2L-1}_{2K_2;2K_1;2K}
\end{displaymath}
\begin{displaymath}
-\Bigl \{
C^{2L_1;2L-1}_{2K_1;2K}I^{2L_2}_{2K_2}-
C^{2L_1;2L-1}_{2K_2;2K}I^{2L_2}_{2K_1}+
C^{2L_1;2L-1}_{2K_2;2K_1}I^{2L_2}_{2K} \Bigr.
\end{displaymath}
\begin{displaymath}
-\Bigl.
C^{2L_2;2L-1}_{2K_1;2K}I^{2L_1}_{2K_2}+
C^{2L_2;2L-1}_{2K_2;2K}I^{2L_1}_{2K_1}-
C^{2L_2;2L-1}_{2K_2;2K_1}I^{2L_1}_{2K} \Bigr \}
\end{displaymath}
\begin{displaymath}
-\Bigl \{
C^{2L_2;2L_1}_{2K_1;2K}I^{2L-1}_{2K_2}-
C^{2L_2;2L_1}_{2K_2;2K}I^{2L-1}_{2K_1}+
C^{2L_2;2L_1}_{2K_2;2K_1}I^{2L-1}_{2K} \Bigr \} =0~,
\end{displaymath}
\begin{displaymath}
(\varepsilon_L+\varepsilon_{L_1}+\varepsilon_{L_2}-
\varepsilon_K-\varepsilon_{K_1}-\varepsilon_{K_2})
C^{2L_2;2L_1;2L-1}_{2K_2-1;2K_1;2K}
\end{displaymath}
\begin{displaymath}
+C^{2L_2;2L_1}_{2K_1;2K}I^{2L-1}_{2K_2-1}+
\Bigl \{
C^{2L_1;2L-1}_{2K_1;2K}I^{2L_2}_{2K_2-1}-
C^{2L_2;2L-1}_{2K_1;2K}I^{2L_1}_{2K_2-1} \Bigr \} =0~,
\end{displaymath}
\begin{displaymath}
(\mu H+\varepsilon_L+\varepsilon_{L_1}+\varepsilon_{L_2}-
\varepsilon_K-\varepsilon_{K_1}-\varepsilon_{K_2})
C^{2L_2;2L_1-1;2L-1}_{2K_2;2K_1;2K-1}
\end{displaymath}
\begin{displaymath}
-\Bigl \{
C^{2L_2;2L-1}_{2K-1;2K_1}I^{2L_1-1}_{2K_2}-
C^{2L_2;2L-1}_{2K-1;2K_2}I^{2L_1-1}_{2K_1}-\Bigr.
\Bigl. 
C^{2L_2;2L_1-1}_{2K-1;2K_1}I^{2L-1}_{2K_2}+
C^{2L_2;2L_1-1}_{2K-1;2K_2}I^{2L-1}_{2K_1}
\Bigr \} 
\end{displaymath}
\begin{displaymath}
-\Bigl \{
C^{2L_1-1;2L-1}_{2K_1;2K-1}I^{2L_2}_{2K_2}-
C^{2L_1-1;2L-1}_{2K_2;2K-1}I^{2L_2}_{2K_1} \Bigr \} =0~,
\end{displaymath}
\begin{displaymath}
(\varepsilon_L+\varepsilon_{L_1}+\varepsilon_{L_2}-
\varepsilon_K-\varepsilon_{K_1}-\varepsilon_{K_2})
C^{2L_2;2L_1-1;2L-1}_{2K_2-1;2K_1;2K-1}
\end{displaymath}
\begin{displaymath}
+\Bigl \{
C^{2L_2;2L-1}_{2K-1;2K_1}I^{2L_1-1}_{2K_2-1}-
C^{2L_2;2L-1}_{2K_2-1;2K_1}I^{2L_1-1}_{2K-1}-\Bigr.
\Bigl. 
C^{2L_2;2L_1-1}_{2K-1;2K_1}I^{2L-1}_{2K_2-1}+
C^{2L_2;2L_1-1}_{2K_2-1;2K_1}I^{2L-1}_{2K-1}
\Bigr \} 
\end{displaymath}
\begin{equation}
+\Bigl \{
C^{2L_1-1;2L-1}_{2K_1;2K-1}I^{2L_2}_{2K_2-1}-
C^{2L_1-1;2L-1}_{2K_1;2K_2-1}I^{2L_2}_{2K-1} \Bigr \} =0~,
\end{equation}
\begin{displaymath}
(\mu H+\varepsilon_L+\varepsilon_{L_1}+\varepsilon_{L_2}-
\varepsilon_K-\varepsilon_{K_1}-\varepsilon_{K_2})
C^{2L_2-1;2L_1-1;2L-1}_{2K_2;2K_1-1;2K-1}
\end{displaymath}
\begin{displaymath}
-\Bigl \{
C^{2L_1-1;2L-1}_{2K_1-1;2K-1}I^{2L_2-1}_{2K_2}-
C^{2L_2-1;2L-1}_{2K_1-1;2K-1}I^{2L_1-1}_{2K_2} + 
C^{2L_2-1;2L_1-1}_{2K_1-1;2K-1}I^{2L-1}_{2K_2}
\Bigr \} =0~,
\end{displaymath}
\begin{displaymath}
(\varepsilon_L+\varepsilon_{L_1}+\varepsilon_{L_2}-
\varepsilon_K-\varepsilon_{K_1}-\varepsilon_{K_2})
C^{2L_2-1;2L_1-1;2L-1}_{2K_2-1;2K_1-1;2K-1}
\end{displaymath}
\begin{displaymath}
+\Bigl \{
C^{2L_1-1;2L-1}_{2K_1-1;2K-1}I^{2L_2-1}_{2K_2-1}-
C^{2L_1-1;2L-1}_{2K_2-1;2K-1}I^{2L_2-1}_{2K_1-1} + 
C^{2L_1-1;2L-1}_{2K_2-1;2K_1-1}I^{2L_2-1}_{2K-1}
\Bigr.
\end{displaymath}
\begin{displaymath}
-C^{2L_2-1;2L-1}_{2K_1-1;2K-1}I^{2L_2-1}_{2K_2-1}+
C^{2L_2-1;2L-1}_{2K_2-1;2K-1}I^{2L_1-1}_{2K_1-1} - 
C^{2L_2-1;2L-1}_{2K_2-1;2K_1-1}I^{2L_1-1}_{2K-1}
\end{displaymath}
\begin{displaymath}
+\Bigl.
C^{2L_2-1;2L_1-1}_{2K_1-1;2K-1}I^{2L-1}_{2K_2-1}-
C^{2L_2-1;2L_1-1}_{2K_2-1;2K-1}I^{2L-1}_{2K_1-1} + 
C^{2L_2-1;2L_1-1}_{2K_2-1;2K_1-1}I^{2L-1}_{2K-1}
\Bigr \} =0~.
\end{displaymath}

Equations (B.1) can easily  be supplemented by scattering 
terms $C^{\cdot\cdot\cdot}_{\cdot\cdot\cdot}\rightarrow 
C^{\cdot\cdot\cdot}_{\cdot\cdot\cdot}$,
and Eqs. (7), (A.1), and (B.1) will still form a 
complete set. 
The structure of interaction Hamiltonian (1) is such that 
scattering leads to connection of the given term only with itself
and with two (or one) neighboring terms. These terms can be 
obtained from the given one by a change of parity of one of the upper or 
lower indexes. The relationships among the various terms 
$C^{\cdot\cdot}_{\cdot\cdot}$ are presented in Fig. 1. 

\newpage
\def\theequation{\thesection.\arabic{equation}}
\setcounter{equation}{0}
\section{Appendix}

We are now able to obtain the self-energy parts 
$\Sigma^{(1)}_{(K,L)}$ and $\Sigma_{(K,L)}$ in fourth-order 
perturbation theory. Straightforward elimination of 
terms in $C^{\cdot\cdot\cdot}_{\cdot\cdot\cdot}$ with $P\geq 2$
from Eqs. (6) using Eqs. (A.2) and (B.1) gives
\begin{displaymath}
\Sigma^{(1)}_{(K,L)}=
\frac{I^{2K_1}_{2L_1}}
{\mu H+\varepsilon_4(L,L_1,K,K_1)-
|I^{2L_2}_{2K_2-1}|^2/
\varepsilon_6-
|I^{2L_2}_{2K_2}|^2/
(\mu H+\varepsilon_6)-\delta E} 
\end{displaymath}
\begin{displaymath}
\times \Biggl \{ I^{2L_1}_{2K_1}-
\frac{I^{2K_2}_{2K_1}}{\mu H+\varepsilon_4(L,L_1,K,K_1)}
\Biggl ( I^{2L_1}_{2K_2}-
\frac{I^{2K_3}_{2K_2}I^{2L_1}_{2K_3}}
{\mu H+\varepsilon_4(L,L_1,K,K_3)}
\end{displaymath}
\begin{displaymath}
+\frac{I^{2L_1}_{2L_2}I^{2L_2}_{2K_2}}
{\mu H+\varepsilon_4(L,L_2,K,K_2)}-
\frac{I^{2K_3-1}_{2K_2}I^{2L_1}_{2K_3-1}}
{\varepsilon_4(L,L_1,K,K_3)} \Biggr )-
\frac{I^{2L_1}_{2L_2}}
{\mu H+\varepsilon_4(L,L_2,K,K_1)}
\end{displaymath}
\begin{displaymath}
\times\Biggl (-I^{2L_2}_{2K_1}+
\frac{I^{2K_2}_{2K_1}I^{2L_2}_{2K_2}}
{\mu H+\varepsilon_4(L,L_2,K,K_2)}-
\frac{I^{2L_2}_{2L_3}I^{2L_3}_{2K_1}}
{\mu H+\varepsilon_4(L,L_3,K,K_1)}+
\frac{I^{2K_2-1}_{2K_1}I^{2L_2}_{2K_2-1}}
{\varepsilon_4(L,L_2,K,K_2)} \Biggr )
\end{displaymath}
\begin{displaymath}
-\frac{I^{2K_2-1}_{2K_1}}
{\varepsilon_4(L,L_1,K,K_2)} 
\Biggl (
I^{2L_1}_{2K_2-1}-
\frac{I^{2K_3}_{2K_2-1}I^{2L_1}_{2K_3}}
{\mu H+\varepsilon_4(L,L_1,K_3,K)}-
\frac{I^{2K_3-1}_{2K_2-1}I^{2L_1}_{2K_3-1}}
{\varepsilon_4(L,L_1,K,K_3)} \Biggr )
\end{displaymath}
\begin{displaymath}
-\frac{I^{2K_2}_{2L_2}}
{\mu H+\varepsilon_6}
\Biggl (
\frac{I^{2L_1}_{2K_2}I^{2L_2}_{2K_1}}{\mu H+\varepsilon_4(L,L_2,K,K_1)}+
\frac{I^{2L_2}_{2K_1}I^{2L_1}_{2K_2}}
{\mu H+\varepsilon_4(L_1,L,K,K_2)}
\end{displaymath}
\begin{displaymath}
-\frac{I^{2L_1}_{2K_1}I^{2L_2}_{2K_2}}
{\mu H+\varepsilon_4(L,L_2,K,K_2)} \Biggr )-
\frac{I^{2K_2-1}_{2L_2}I^{2L_1}_{2K_2-1}I^{2L_2}_{2K_1}}
{\varepsilon_6(\mu H+\varepsilon_4(L,L_2,K,K_1)} \Biggr \}
\end{displaymath}
\begin{equation}
+\frac{I^{2K_1-1}_{2L_1}}
{\varepsilon_4(L,L_1,K,K_1)-
|I^{2K_2}_{2L_2-1}|^2/
(\mu H+\varepsilon_6)-
|I^{2K_2-1}_{2L_2-1}|^2/
\varepsilon_6-\delta E} 
\end{equation}
\begin{displaymath}
\times\Biggl \{ I^{2L_1}_{2K_1-1}-
\frac{I^{2K_2}_{2K_1-1}}{\mu H+\varepsilon_4(L,L_1,K,K_2)}
\Biggl ( I^{2L_1}_{2K_2}-
\frac{I^{2K_3}_{2K_2}I^{2L_1}_{2K_3}}
{\mu H+\varepsilon_4(L,L_1,K,K_3)}
\end{displaymath}
\begin{displaymath}
+\frac{I^{2L_1}_{2L_2}I^{2L_2}_{2K_2}}
{\mu H+\varepsilon_4(L,L_2,K,K_2)}-
\frac{I^{2K_3-1}_{2K_2}I^{2L_1}_{2K_3-1}}
{\varepsilon_4(L,L_1,K,K_3)} \Biggr )-
\frac{I^{2K_2-1}_{2K_1-1}}
{\varepsilon_4(L,L_1,K,K_2)}
\end{displaymath}
\begin{displaymath}
\times\Biggl (I^{2L_1}_{2K_2-1}-
\frac{I^{2K_3}_{2K_2-1}I^{2L_1}_{2K_3}}
{\mu H+\varepsilon_4(L,L_1,K,K_3)}-
\frac{I^{2K_3-1}_{2K_2-1}I^{2L_1}_{2K_3-1}}
{\varepsilon_4(L,L_1,K,K_3)} \Biggr )
\end{displaymath}
\begin{displaymath}
+\frac{I^{2L-1}_{2L_2-1}I^{2L_2-1}_{2L_3}I^{2L_3}_{2K_1-1}}
{(\mu H+\varepsilon_4(L,L_2,K,K_1))
\varepsilon_4(L,L_3,K,K_1)}- 
\frac{I^{2K_2-1}_{2L_2-1}I^{2L_2-1}_{2K_1-1}I^{2L_1}_{2K_2-1}}
{\varepsilon_6\varepsilon_4(L,L_1,K,K_2)} \Biggr \}~,
\end{displaymath}
\begin{displaymath}
\Sigma_{(K,L)}=
\frac{I^{2K_1}_{2L_1-1}}
{\mu H+\varepsilon_4(L,L_1,K,K_1)-\delta E-
|I^{2L_2}_{2K_2-1}|^2/
\varepsilon_6-
|I^{2L_2}_{2K_2}|^2/
(\mu H+\varepsilon_6)}
\end{displaymath}
\begin{displaymath}
\times\Biggl \{ I^{2L_1-1}_{2K_1}-
\frac{I^{2K_2}_{2K_1}}{\mu H+\varepsilon_4(L,L_1,K,K_2)}
\Biggl ( I^{2L_1-1}_{2K_2}-
\frac{I^{2K_3}_{2K_2}I^{2L_1-1}_{2K_3}}
{\mu H+\varepsilon_4(L,L_1,K,K_3)}
\end{displaymath}
\begin{displaymath}
-\frac{I^{2K_3-1}_{2K_2}I^{2L_1-1}_{2K_3-1}}
{\varepsilon_4(L,L_1,K,K_3)} \Biggr )-
\frac{I^{2K_2-1}_{2K_1}}
{\varepsilon_4(L,L_1,K,K_2)} 
\Biggl ( I^{2L_1-1}_{2K_2-1}-
\frac{I^{2K_3}_{2K_2-1}I^{2L_1-1}_{2K_3}}
{\mu H+\varepsilon_4(L,L_1,K,K_3)}
\end{displaymath}
\begin{displaymath}
+\frac{I^{2L_1-1}_{2L_2-1}I^{2L_2-1}_{2K_2-1}}
{\varepsilon_4(L,L_2,K,K_2)}-
\frac{I^{2K_3-1}_{2K_2-1}I^{2L_1-1}_{2K_3-1}}
{\varepsilon_4(L,L_1,K,K_3)} \Biggr )+
\frac{I^{2L_1-1}_{2L_2}I^{2L_2}_{2L_3-1}I^{2L_3-1}_{2K_1}}
{\varepsilon_4(L,L_2,K,K_1)(\mu H+\varepsilon_4(L,L_3,K,K_1))} 
\end{displaymath}
\begin{equation}
-\frac{I^{2K_2}_{2L_2}I^{2L_2}_{2K_1}I^{2L_1-1}_{2K_2}}
{(\mu H+\varepsilon_6)(\mu H+\varepsilon_4(L,L_1,K,K_2))} 
\Biggr \} 
\end{equation}
\begin{displaymath}
+\frac{I^{2K_1-1}_{2L_1-1}}
{\varepsilon_4(L,L_1,K,K_1)-
|I^{2K_2}_{2L_2-1}|^2 \bigl /
(\mu H+\varepsilon_6) -
|I^{2K_2-1}_{2L_2-1}|^2 \bigl /
\varepsilon_6-\delta E} 
\end{displaymath}
\begin{displaymath}
\times\Biggl \{
I^{2L_1-1}_{2K_1-1}-
\frac{I^{2K_2}_{2K_1-1}}{\mu H+\varepsilon_4(L,L_1,K,K_2)}
\Biggl ( I^{2L_1-1}_{2K_2}-
\frac{I^{2K_3}_{2K_2}I^{2L_1-1}_{2K_3}}
{\mu H+\varepsilon_4(L,L_1,K,K_3)}
\end{displaymath}
\begin{displaymath}
-\frac{I^{2K_3-1}_{2K_2}I^{2L_1-1}_{2K_3-1}}
{\varepsilon_4(L,L_1,K,K_3)} \Biggr )-
\frac{I^{2K_2-1}_{2K_1-1}}
{\varepsilon_4(L,L_1,K,K_2)} \Biggl ( I^{2L_1-1}_{2K_2-1}
\end{displaymath}
\begin{displaymath}
-\frac{I^{2K_3}_{2K_2-1}I^{2L_1-1}_{2K_3}}
{\mu H+\varepsilon_4(L,L_1,K,K_3)} +
\frac{I^{2L_1-1}_{2L_2-1}I^{2L_2-1}_{2K_2-1}}
{\varepsilon_4(L,L_2,K,K_2)}-
\frac{I^{2K_3-1}_{2K_2-1}I^{2L_1-1}_{2K_3-1}}
{\varepsilon_4(L,L_1,K,K_3)} \Biggr ) 
\end{displaymath}
\begin{displaymath}
+\frac{I^{2L_1-1}_{2L_2-1}}
{\varepsilon_4(L,L_2,K,K_1)}
\Biggl ( I^{2L_2-1}_{2K_1-1}-
\frac{I^{2K_2}_{2K_1-1}I^{2L_2-1}_{2K_2}}
{\mu H+\varepsilon_4(L,L_2,K,K_2)}
\end{displaymath}
\begin{displaymath}
+\frac{I^{2L_2-1}_{2L_3-1}I^{2L_3-1}_{2K_1-1}}
{\varepsilon_4(L,L_3,K,K_1)}-
\frac{I^{2K_2-1}_{2K_1-1}I^{2L_2-1}_{2K_2-1}}
{\varepsilon_4(L,L_2,K,K_2)} \Biggr )+
\frac{I^{2K_2-1}_{2L_2-1}}
{\varepsilon_6}
\end{displaymath}
\begin{displaymath}
\times\Biggl (
\frac{I^{2L_1-1}_{2K_1-1}I^{2L_2-1}_{2K_2-1}}
{\varepsilon_4(L,L_2,K,K_2)}-
\frac{I^{2L_1-1}_{2K_2-1}I^{2L_2-1}_{2K_1-1}}
{\varepsilon_4(L,L_2,K,K_1)} -
\frac{I^{2L_2-1}_{2K_1-1}I^{2L_1-1}_{2K_2-1}}
{\varepsilon_4(L,L_1,K,K_2)} \Biggr )
\end{displaymath}
\begin{displaymath}
-\frac{I^{2K_2}_{2L_2-1}I^{2L_1-1}_{2K_2}I^{2L_2-1}_{2K_1-1}}
{(\mu H+\varepsilon_6)\varepsilon_4(L,L_2,K,K_1)} 
\Biggr \}.
\end{displaymath}

From Eqs. (6) and (A.2), the quantities $C^{2L-1}_{2K}$ and 
$C^{2L-1}_{2K-1}$ can easily be obtained in the third 
order of perturbation 
theory. We do not give these expressions here because only 
one statement is essential for us: direct comparison of the 
quantities $\delta E$ (Eq. (4)) and self-energy $\Sigma_{K,L}$ 
(Eq. (C.2)) shows that 
\begin{equation}
\delta E+\Sigma (K,L)_{\varepsilon_K=\varepsilon_L=\varepsilon_F}=0.
\end{equation}
Equation (C.3) is valid for arbitrary spectrum 
$\varepsilon_K, \varepsilon_L$ and arbitrary transition matrix 
elements $I^{\cdot}_{\cdot}$. Our  conjecture is that Eq. (C.3) holds 
in all orders of perturbation theory, and hence we can put
\begin{equation}
\delta E +\Sigma (K,L)_{\varepsilon_K=\varepsilon_L=\varepsilon_F}=
-\Delta,
\end{equation}
where $\Delta$ is exponentially small and can be considered  an
order parameter. We also obtain from Eqs. (C.1) and (C.2) that self-energies 
$\Sigma^{(1)}$ and $\Sigma$  coincide only in the second order of 
perturbation theory. They start to be different in the third order 
of perturbation theory. In the fourth order of perturbation theory, we 
obtain from Eqs. (C.1) and (C.2) 
\begin{displaymath}
\Sigma^{(1)}_{(K,L)}-\Sigma_{(K,L)}=
I^{2K_1}_{2L_1}I^{2L_1}_{2L_2}I^{2L_2}_{2K_1}
\Biggl (
\frac{1}{(\mu H+\varepsilon_4(L,L_1,K,K_1)
(\mu H+\varepsilon_4(L,L_2,K,K_1))}
\end{displaymath}
\begin{displaymath}
-\frac{1}{\varepsilon_4(L,L_1,K,K_1)\varepsilon_4(L,L_2,K,K_1)}
\Biggr ) -I^{2K_2}_{2K_1}I^{2K_1}_{2L_1}I^{2L_1}_{2L_2}I^{2L_2}_{2K_2}
\end{displaymath}
\begin{equation}
\times\Biggl \{  \Biggr.
\Biggl (\frac{1}{\mu H+\varepsilon_4(L,L_1,K,K_1)}+
\frac{1}{\varepsilon_4(L,L_1,K,K_1)} \Biggr ) 
\end{equation}
\begin{displaymath}
\times\Biggl (
\frac{1}{(\mu H+\varepsilon_4(L,L_2,K,K_2))
(\mu H+\varepsilon_4(L,L_1,K,K_2))}
\end{displaymath}
\begin{displaymath}
-\frac{1}{\varepsilon_4(L,L_1,K,K_2)
\varepsilon_4(L,L_2,K,K_2)} \Biggr )+
\Biggl (\frac{1}{\mu H+\varepsilon_4(L,L_2,K,K_2)}
\end{displaymath}
\begin{displaymath}
+\frac{1}{\varepsilon_4(L,L_2,K,K_2)} \Biggr ) 
\Biggl (\frac{1}{(\mu H+\varepsilon_4(L,L_2,K,K_1))
(\mu H+\varepsilon_4(L,L_1,K,K_1))}
\end{displaymath}
\begin{displaymath}
-\frac{1}{\varepsilon_4(L,L_2,K,K_1)\varepsilon_4(L,L_1,K,K_1)}
\Biggr )-
\Biggl ( \frac{1}
{\mu H+\varepsilon_4(L,L_2,K,K_1)}-
\frac{1}{\varepsilon_4(L,L_2,K,K_1)} \Biggr )
\end{displaymath}
\begin{displaymath}
\times\Biggl (
\frac{1}
{(\mu H+\varepsilon_4(L,L_3,K,K_1))
(\mu H+\varepsilon_4(L,L_1,K,K_1)}
\end{displaymath}
\begin{displaymath}
+\frac{1}
{\varepsilon_4(L,L_3,K,K_1)
\varepsilon_4(L,L_1,K,K_1)} \Biggr )
\end{displaymath}
\begin{displaymath}
+\Biggl (
\frac{1}
{\varepsilon_6(\mu H+\varepsilon_4(L,L_2,K,K_1))
(\mu H+\varepsilon_4(L,L_1,K,K_1))} 
\end{displaymath}
\begin{displaymath}
-\frac{1}
{(\mu H+\varepsilon_6)\varepsilon_4(L,L_2,K,K_1)
\varepsilon_4(L,L_1,K,K_1)} \Biggr ) 
\end{displaymath}
\begin{displaymath}
+\Biggl (
\frac{1}{(\mu H+\varepsilon_6)(\mu H+\varepsilon_4(L,L_1,K,K_1)} 
\Biggr ) \cdot
\Biggl ( \frac{1}
{\mu H+\varepsilon_4(L,L_2,K,K_1)}
\end{displaymath}
\begin{displaymath}
-\frac{1}{\mu H+\varepsilon_4(L,L_2,K,K_2)} \Biggr )-
\Biggl ( \frac{1}
{\varepsilon_4(L,L_2,K,K_1)}-
\frac{1}{\varepsilon_4(L,L_2,K,K_2)} \Biggr ) 
\end{displaymath}
\begin{displaymath}
\times\frac{1}{\varepsilon_6\varepsilon_4(L,L_1,K,K_1)} \Biggr \},
\end{displaymath}
where
\begin{equation}
\varepsilon_4(L,L_1,K,K_1)\equiv\varepsilon_L+\varepsilon_{L_1}-
\varepsilon_K-\varepsilon_{K_1}~,
\end{equation}
\begin{displaymath}
\varepsilon_6\equiv\varepsilon_L+\varepsilon_{L_1}+\varepsilon_{L_2}-
\varepsilon_K-\varepsilon_{K_1}-\varepsilon_{K_2}.
\end{displaymath}
Straightforward calculation of the integrals in Eq. (C.5) leads to 
Eqs. (40) and (41). Both Eqs. (40) and (41) are proved in two 
orders of perturbation theory. Our conjecture is that 
Eq. (41) is exact.

\newpage
\def\theequation{\thesection.\arabic{equation}}
\setcounter{equation}{0}
\section{Appendix }

In this appendix we consider the role of the right-hand side of Eqs. (7)
for a negative value of the coupling constant, $g<0$. 
In the first order of perturbation theory, we obtain from (A.2)
\begin{displaymath}
C^{2L_1;2L-1}_{2K_1;2K}=
\frac{1}{\mu\tilde H+\varepsilon_4(L,L_1,K,K_1)+\Delta}
\end{displaymath}
\begin{displaymath}
\times\Biggl [ C^{2L-1}_{2K_1}I^{2L_1}_{2K}-C^{2L-1}_{2K}I^{2L_1}_{2K_1}+
C^{2L_1}_{2K}I^{2L-1}_{2K_1}-C^{2L_1}_{2K_1}I^{2L-1}_{2K} \Biggr ]~;
\end{displaymath}
\begin{equation}
C^{2L_1;2L-1}_{2K_1-1;2K}=
\frac{1}{\varepsilon_4(L,L_1,K,K_1)+\Delta}
\Biggl [ 
I^{2L_1}_{2K_1-1}C^{2L-1}_{2K}-C^{2L_1}_{2K}I^{2L-1}_{2K_1-1}
\Biggr ]~;
\end{equation}
\begin{displaymath}
C^{2L_1-1;2L-1}_{2K_1;2K-1}=
\frac{1}{\mu\tilde H+\varepsilon_4(L,L_1,K,K_1)+\Delta}
\Biggl [ C^{2L_1-1}_{2K-1}I^{2L-1}_{2K_1}-C^{2L-1}_{2K-1}I^{2L_1-1}_{2K_1}
\Biggr ]~;
\end{displaymath}
\begin{displaymath}
C^{2L_1-1;2L-1}_{2K_1-1;2K-1}=
\frac{1}{\varepsilon_4(L,L_1,K,K_1)+\Delta}
\end{displaymath}
\begin{displaymath}
\times\Biggl [ 
I^{2L_1-1}_{2K_1-1}C^{2L-1}_{2K-1}-C^{2L-1}_{2K_1-1}I^{2L_1-1}_{2K-1}+
C^{2L_1-1}_{2K_1-1}I^{2L-1}_{2K-1}-C^{2L_1-1}_{2K-1}I^{2L-1}_{2K_1-1}
\Biggr ].
\end{displaymath}
Inserting (D.1) into (6), we obtain
\begin{displaymath}
A_1\Bigl ( C^{2L-1}_{2K};C^{2L-1}_{2K-1},C^{2L}_{2K} \Bigr )=-
\sum \frac
{I^{2K_1}_{2L_1}\Bigl ( 
C^{2L-1}_{2K_1}I^{2L_1}_{2K}+
C^{2L_1}_{2K}I^{2L-1}_{2K_1}-
C^{2L_1}_{2K_1}I^{2L-1}_{2K} \Bigr )}
{\mu\tilde H+\varepsilon_4(L,L_1,K,K_1)+\Delta}
\end{displaymath}
\begin{equation}
-\sum \frac
{I^{2K_1-1}_{2L_1}I^{2L-1}_{2K_1-1}C^{2L_1}_{2K}}
{\varepsilon_4(L,L_1,K,K_1)+\Delta}~,
\end{equation}
\begin{displaymath}
A_2\Bigl ( C^{2L-1}_{2K};C^{2L-1}_{2K-1},C^{2L}_{2K} \Bigr )=
\end{displaymath}
\begin{displaymath}
-\sum \frac
{I^{2K_1-1}_{2L_1-1}\Bigl ( 
C^{2L-1}_{2K_1-1}I^{2L_1-1}_{2K-1}-
C^{2L_1-1}_{2K_1-1}I^{2L-1}_{2K-1}+
C^{2L_1-1}_{2K-1}I^{2L-1}_{2K_1-1} \Bigr )}
{\varepsilon_4(L,L_1,K,K_1)+\Delta}
\end{displaymath}
\begin{displaymath}
-\sum \frac
{I^{2K_1}_{2L_1-1}I^{2L-1}_{2K_1}C^{2L_1-1}_{2K-1}}
{\mu\tilde H+\varepsilon_4(L,L_1,K,K_1)+\Delta}~,
\end{displaymath}
\begin{displaymath}
A_3\Bigl ( C^{2L-1}_{2K};C^{2L-1}_{2K-1},C^{2L}_{2K} \Bigr )=
\sum \frac
{I^{2K_1}_{2L_1-1}\Bigl ( 
C^{2L_1-1}_{2K_1}I^{2L}_{2K}-
C^{2L_1-1}_{2K}I^{2L}_{2K_1}-
C^{2L}_{2K_1}I^{2L_1-1}_{2K} \Bigr )}
{\mu\tilde H+\varepsilon_4(L,L_1,K,K_1)+\Delta}
\end{displaymath}
\begin{displaymath}
-\sum \frac
{I^{2K_1-1}_{2L_1-1}I^{2L}_{2K_1-1}C^{2L_1-1}_{2K}}
{\varepsilon_4(L,L_1,K,K_1)+\Delta}
\end{displaymath}
The quantities $\varepsilon_4,~\varepsilon_6$ here are the same as in
Eq. (C.6). 

As before, only convolutions $Z_L,Y_L$ are large for $g<0$. 
Furthermore,  
\begin{equation}
|Z_L+Y_L|\sim g^2|Z_L-Y_L|.
\end{equation}
As the result, Eqs. (7) can be reduced to just one equation:  
\begin{equation}
\Bigl ( Z_L-Y_L \Bigr )
\Biggl [ 1+g\ln\frac{\varepsilon_F}{y+\Delta}+
g\ln\frac{\varepsilon_F}{\mu\tilde H+y+\Delta}
\end{equation}
\begin{displaymath}
+\frac{g^3}{2}\Biggl (
\frac{I_1}{g\ln\varepsilon_F \bigl /(\mu\tilde H+y+\Delta)}+
\frac{I_2}{g\ln\varepsilon_F \bigl /(y+\Delta)} \Biggr ) \Biggr ] 
\end{displaymath}
\begin{displaymath}
=Ig \Biggl ( \ln\frac
{\varepsilon_f}{\mu\tilde H+y+\Delta}+
\ln \frac{\varepsilon_F}{y+\Delta} \Biggr )+
g\int dx \Biggl (
\frac{X_K}{\mu\tilde H+y+x+\Delta}-
\frac{Y_K}{y+x+\Delta} \Biggr ),
\end{displaymath}
where
\begin{equation}
I_1=\int\frac{dxdydx_1}
{(\mu\tilde H+y+x+\Delta)
(\mu\tilde H+y+x_1+\Delta)
(\mu\tilde H+y+x+y_1+x_1+\Delta)},
\end{equation}
\begin{displaymath}
I_2=\int\frac{dxdydx_1}
{(y+x+\Delta)
(y+x_1+\Delta)
(y+x+y_1+x_1+\Delta)}.
\end{displaymath}
A simple calculation of the integrals (D.5) gives
\begin{equation}
I_1=\frac{1}{3}\ln^3\Biggl (
\frac{\varepsilon_F}{\mu\tilde H+y+\Delta} \Biggr )~, \qquad
\end{equation}
\begin{displaymath}
I_2=\frac{1}{3}\ln^3\Biggl (
\frac{\varepsilon_F}{y+\Delta} \Biggr ).
\end{displaymath}
Now we can define the Kondo temperature $T_c$ to be 
\begin{equation}
|g|\ln \frac{\varepsilon_F}{T_c}=z,
\end{equation}
where $z$ is a root of the ???quadratic equation
\begin{equation}
1-2z+\frac{z^2}{3}=0; \qquad
z=3-\sqrt{6}\approx 0.5505.
\end{equation}
From Eq. (D.4) we obtain
\begin{equation}
Z_L-Y_L=-
\frac{I\tilde\beta}
{|g|(1-z/3)\ln
\Biggl ( \frac
{(\mu\tilde H+y+\Delta)(y+\Delta)}{T^2_c}\Biggr )},
\end{equation}
where $\tilde\beta$ is a number of order  1. Instead of Eqs. 
(41) and (42) we have now
\begin{equation}
\mu\tilde H=\mu H-\delta\Sigma; \qquad
\delta \Sigma=-\mu Hz(-1/2+\langle S_z\rangle ).
\end{equation}
As before, the average spin $\langle S_z\rangle$ is given by 
Eq. (27) with the replacement $\mu H\rightarrow \mu\tilde H$:
\begin{equation}
\langle S_z\rangle =
\frac{\mu\tilde H}{4\bigl (T^2_c+(\mu\tilde H/2)^2\bigr )^{1/2}}.
\end{equation}

The magnetic field dependence of 
the average spin $\langle S_z\rangle$ (Eqs. (D.10) and  (D.11)) 
is given in Fig. 2. Dots are the experimental results of Ref. 4.


\newpage
\begin{thebibliography}{99}
\bibitem{1} A.~A.~Abrikosov and A.~A.~Migdal, 
J. of Low Temp. Phys. {\bf 3}, 519 (1970).
\bibitem{2}A.~M.~Tsvelick and P.~B.~Wigmann, Advances in 
Physics {\bf 32}, 453 (1983).
\bibitem{3} N.~Andrei, K.~Furuya, and J.~H.~Lowenstein, 
Rev. Mod. Phys. {\bf 55}, 331 (1983).
\bibitem{4} W.~Felsch, Z. Phys.
{\bf B29}, 212 (1978).
\bibitem{5} S.~D.~Bader, N.~E.~Phillips, M.~B.~Haple, and 
C.~A.~Luengo, Solid State Commun. {\bf 16}, 1263 (1975).
\bibitem{6} P.~Nozieres and T.C. de Dominicis, 
Phys. Rev. {\bf 178}, 1097 (1969).
\bibitem{7} Yu.~N.~Ovchinnikov, A.~M.~Dyugaev,  P.~Fulde,
and V.~Z.~Kresin, JETP Lett. {\bf 66}, 184 (1997).
\bibitem{8} K.~Yosida, Phys. Rev. {\bf 147}, 223 (1966).
\bibitem{9} Hiroumi Ishii, Prog. Theor. Phys.
{\bf 40}, 201 (1968).
\bibitem{10} Hiroumi Ishii, Prog. Theor. Phys. {\bf 43},
578 (1970).
\bibitem{11} M.~Fowler, A.~Zawadowskii,
Solid St. Comm. {\bf 9}, 471 (1971). 
\end{thebibliography}
\end{document}